\RequirePackage{rotating}
\RequirePackage{lineno}
\PassOptionsToPackage{usenames,dvipsnames}{xcolour}

\documentclass[a4paper,twocolumn, numberedappendix, onecolappendix,revtex4,iop]{openjournal} 
\usepackage[a4paper,marginpar=10pt,headheight=20pt, right=1cm, top=1.5in, left=1in, bottom=0pt, footskip=0pt, footnotesep=10pt]{geometry}

\pdfoutput=1 
\usepackage{amsmath,amstext}
\usepackage[T1]{fontenc}
\usepackage{apjfonts}
\usepackage{ae,aecompl}
\usepackage[utf8]{inputenc}
\usepackage{natbib}
\usepackage{url}
\usepackage{mdwlist}
\usepackage{listings}
\usepackage{rotating}
\usepackage{multirow}
\usepackage{times}
\newcommand{\RNum}[1]{\uppercase\expandafter{\romannumeral #1\relax}}

\usepackage{booktabs}
\usepackage{multirow}
\usepackage{placeins}

\usepackage[colorlinks=true,linkcolor=blue,citecolor=blue,urlcolor=blue ]{hyperref}
\usepackage{xurl} 
\usepackage{xspace}
\usepackage{pifont}
\usepackage{orcidlink}
\usepackage{xcolor}          

\usepackage{ulem}
\usepackage{comment}

\newcommand{\eg}{\textit{e.g.}}
\newcommand{\ie}{\textit{i.e.}}

\shorttitle{Exploring the Evidence for Evolving Dark Energy with the Jeffreys prior}
\shortauthors{}

\begin{document}
\title{Exploring the Evidence for Evolving Dark Energy with the Jeffreys prior}

\author{\vspace*{-35px}
    Joshua Layton$^{1\star}$,
    Alex Hall$^{1}$
    and
    Naomi Clare Robertson$^{1}$
    }
\thanks{$^\star$ \href{mailto:email}{s2161585@ed.ac.uk}}
\affiliation{
    $^{1}$Institute for Astronomy, University of Edinburgh, Royal Observatory, Blackford Hill, Edinburgh EH9\,3HJ, UK
    }

\begin{abstract}
We investigate the impact of applying the uninformative Jeffreys prior in a re-analysis of DESI DR2 BAO, Planck 2018 CMB, and Type Ia Supernovae (DES-SN5YR, DES-Dovekie, Pantheon+, Union3). We incorporate this prior into new posterior inference and evidence calculations. We introduce novel techniques for interpreting the Jeffreys prior physically. To overcome the computational expense of the Jeffreys prior, we develop a fully \texttt{JAX}-based pipeline consisting of \texttt{CosmoPower-JAX} emulators, gradient-based posterior inference using the No-U-Turn Sampler (NUTS), and evidence calculations using \texttt{JAX}-accelerated nested sampling. We demonstrate that while the parameter constraints on \(w_0\) and \(w_a\) remain robust to prior choice (exhibiting only minor shifts of the marginalised posterior towards the \textit{Maximum A Posteriori} in less constraining dataset combinations), the Bayesian evidence is highly sensitive to the use of the Jeffreys prior; replacing uniform priors with the Jeffreys prior reverses the model-selection preference in dataset combinations that favour dynamical dark energy under the uniform prior, while strengthening existing preferences for \(\Lambda\)CDM. This is driven primarily by the significantly larger Occam penalty applied by the Jeffreys prior which penalises the overfitting of the data by a \(w_0w_a\)CDM model.
\end{abstract}

\maketitle

\section{Introduction}\label{sec:introduction}
The analysis of DESI DR1 BAO data by the DESI Collaboration \citep{adame2025desi} made the profound statement that the widely accepted \(\Lambda\)CDM model is disfavoured when compared  to the dynamical dark energy model, \(w_0w_a\)CDM \citep{chevallier2001accelerating,linder2003exploring}. This evidence persisted in the more recent DESI Collaboration analysis of BAO in DESI DR2 \citep{abdul2025desi}, in which the preference for \(w_0w_a\)CDM reached the 4.2\(\sigma\) level when considering the BAO data in conjunction with the Planck cosmic microwave background (CMB) and DES-SN5YR supernovae (SN). Recent studies have investigated these findings using Bayesian model selection \citep{ong2026bayesian,hergt2026consistency}, which suggest that the preference for dynamical dark energy only holds under particular SN data. An especially important case is the DES-SN5YR reanalysis, DES-Dovekie \citep{popovic2026dark}, which reduces the evidence in favour of \(w_0w_a\)CDM, but still holds at \(5:1\) posterior odds. A recent paper has also produced a new SN sample, called the `Unite' sample \citep{camilleri2026supernovae}, which combines the Pantheon+ and DES-SN5YR samples. This sample follows a similar trend that indicates weak evidence for evolving dark energy using Bayesian evidence and a \(3.1\sigma\) significance using the maximum likelihood frequentist analysis.

The challenges which arise in adopting the Bayesian framework are the sensitivity to prior choice and susceptibility to prior volume effects. Prior volume effects or projection effects, are the misalignment between the marginalised posterior mode and the \textit{maximum a posteriori} (MAP) estimate caused by integrating over large anisotropic nuisance volumes. Such effects can cause marginalised posteriors to shift away from the true parameter constraints \citep{holm2023bayesian,raveri2024understanding}; fortunately studies such as \cite{tsedrik2025simple} and \cite{bonici2026alleviating} have shown that reparametrisation of nuisance parameters is an effective way of mitigating prior volume effects while remaining easy to implement. Projection effects do become increasingly prevalent when we adopt wide, supposedly `uninformative', flat priors. Uniform priors are generally adopted to show ignorance in parameters; however, this notion of being uninformative can be misleading. While we can have a uniform prior in one space, when transformed to an alternative space, the prior can become highly informative, which gives rise to reparametrisation bias. This raises the question of whether the posterior constraints obtained are a consequence of the idiosyncratic parametrisation used. Additionally, when we study model comparison, we find that the Bayesian evidence is highly sensitive to the prior we choose; for example, decreasing the lower bound on \(w_0\) and \(w_a\) can significantly impact conclusions about the evidence for dynamical dark energy \citep{patel2024prior}.

Unlike other fields of physics, where the imposition of a prior is typically motivated by physical reasoning, favoured dark energy models are phenomenological in nature. Hence, priors on such models are physically arbitrary, unless a specific microphysical model is assumed to inform prior choices (\eg, \citealt{toomey2025theory}). Such subjectivity makes model selection conclusions sensitive to arbitrary choices; therefore, it is advantageous to investigate the robustness of claims under alternative prior constructions. The most pertinent claim of this work is that if the tentative evidence for \(w_0w_a\) was confirmed, it would have very significant implications for cosmology and fundamental physics. It is therefore important to scrutinise the statistical evidence and check its robustness to underlying assumptions such as the adopted priors. Thus, we consider the reliability of the DESI Collaboration dynamical dark energy results under a statistically motivated prior, the Jeffreys prior \citep{jeffreys1946invariant}, which is known to be the theoretically least informative prior in one dimension, and invariant under reparametrisation. 

The Jeffreys Prior is of particular interest because it is less informative than the commonly assumed uniform prior for most parameters in a typical cosmological analysis. By re-analysing the evidence for dynamical dark energy with the Jeffreys prior, we can perform the most conservative Bayesian analysis possible within the \(w_0w_a\)CDM model. While this is clearly a valuable test of the evidence for/against \(\Lambda\)CDM, there are significant technical challenges in implementing the Jeffreys prior, as it requires likelihood derivatives at every point in parameter space. In this work, we overcome these challenges with a new \texttt{JAX}-based analysis pipeline\footnote{\url{https://github.com/JoshL965/Jeffreys-prior-project}}. We consider the effects of introducing the Jeffreys prior on the dataset combinations, BAO + CMB and BAO + CMB + SN, along with extensions to the base \(w_0w_a\)CDM model (free curvature and free neutrino mass), with a particular focus on the DES-SN5YR SN data, due to it being the strongest supporter of a \(w_0w_a\)CDM Universe.

The rest of this paper is laid out as follows: Section \ref{sec:datasets} briefly covers the datasets we use in this work. In Section \ref{sec:methodology}, we start by covering the Jeffreys prior, how it is implemented, and the physical intuition behind its behaviour. We then proceed to the cosmological posterior inference pipeline we develop, the priors we set, and the settings we apply. This then leads into the training of the emulators used for theoretical predictions, and finally the method used for obtaining the Bayesian evidence. Then in Section \ref{sec:results} we present our results, the comparison of posteriors and Bayes factors, and provide a discussion of their implications. Finally, in Section \ref{sec:conclusion}, we summarise our findings and suggest future improvements to our analysis.

\section{Datasets}\label{sec:datasets}
In this work we follow closely the approach taken in the DESI Collaboration analysis presented in \cite{adame2025desi}. As such, the datasets used here go as follows: For BAO data, we use the DESI Collaborations latest release, DESI DR2 \citep{abdul2025desi}, which provides the most precise measurements to date, covering a redshift range, \(0.1 < z < 3.5\). Considering Type Ia supernovae data (labelled SN hereafter) we analyse a range of datasets, including Pantheon+ (PantheonPlus) \citep{brout2022pantheon+}, Union3 \citep{rubin2025union}, the DES reanalysis, DES-Dovekie (Dovekie) \citep{popovic2026dark}, and a particular emphasis on the original analysis, DES-SN5YR (DESY5) \citep{abbott2024dark}.
Our implementation of the Jeffreys prior assumes that all likelihoods are Gaussian. This is the case for BAO, SN and high-\(\ell\) CMB data, but is not the case for low-\(\ell\) CMB data. Thus in this work we choose to adopt the compressed low-\(\ell\) TT and EE likelihoods supplied by \cite{prince2022compressed} in \texttt{Planck-low-py}. These binned likelihoods are well approximated by log-normal distributions which under transformation satisfy the Gaussian requirement in the Jeffreys prior implementation. The high-\(\ell\) TT, TE, and EE likelihoods within \texttt{Planck-low-py} are consistent with \texttt{plik-lite} employed by the DESI Collaboration in \cite{adame2025desi}. In contrast to the DESI Collaboration analysis, we do not implement the CMB lensing likelihood, our main motivation being that it will not impact our results significantly, and we wish to keep our analysis as simple as possible.

\section{Methodology}\label{sec:methodology}
\subsection{Jeffreys prior}\label{sec:jeffreysprior}
The theoretically least informative prior in one dimension takes the form of the Jeffreys prior \citep{jeffreys1946invariant}. Given some parameter vector \(\theta\), the Jeffreys prior is given by 
\begin{equation}
    \pi_\text{J}(\theta) \propto \sqrt{|\mathcal{F}(\theta)|}.
\end{equation}
Where \(\mathcal{F(\theta})\) is the Fisher information matrix (FIM). The Jeffreys prior has the appealing quality of invariance under reparametrisation which manifests from the FIM transforming as a tensor \citep{bonici2026alleviating}. This removes any unwanted bias potentially introduced in the marginalised posterior caused by the choice of parameter space.

The Jeffreys prior has also been shown to reduce prior volume effects \citep{tsedrik2025simple,bonici2026alleviating}. The act of integrating over large anisotropic volumes in the unmarginalised posterior causes low probability density regions to influence the marginalised posterior, shifting the marginalised posterior mode away from the \textit{maximum a posteriori} (MAP) \citep{raveri2024understanding}. As the FIM measures the curvature of the log-likelihood, large, flat, low probability density regions are penalised by the Jeffreys prior, hence minimising some prior-volume effects. Specifically, volume effects present from nuisance parameters which appear linearly in a model are fully alleviated by the Jeffreys prior due to the cancellation of the Laplace term in the unmarginalised posterior distribution \citep{tsedrik2025simple}.

In more than one dimension, the Jeffreys prior is not without its pathologies. First of all, as noted by \cite{kass1996selection}, the multidimensional Jeffreys prior can be overly aggressive and apply excessive weight to specific regions of the parameter space. In the cosmological context, this manifests as a severe prior preference for the highly distinguishable extremes of a model. Secondly, applying the full joint prior is susceptible to marginalisation paradoxes, whereby when integrating over unconstrained nuisance volumes, the prior on the target parameters can become pathologically dominated by the geometry of the nuisance space \citep{dawid1973marginalization}. Other pathologies do exist; however, most arise due to the improper nature of the Jeffreys prior, such as the marginalisation paradox. In our analysis, we require the Jeffreys prior to be proper; hence, we truncate the Jeffreys prior (see Section \ref{sec:CosmoInfer}), therefore, avoiding many of these pathologies. Despite these theoretical limitations, our decision to implement the Jeffreys prior in parameter inference and Bayesian model selection remains well grounded. In the high S/N limit of BAO, CMB, and SN measurements, the highly peaked likelihood should overpower the overly aggressive prior while allowing the Jeffreys prior to remain objective to parametrisation and potentially mitigate projection effects. It is also important to reiterate that the purpose of this work is not to prove the Jeffreys prior to be uniquely favoured, but instead to test how robust the claims of dynamical dark energy are under an alternative, statistically well-motivated, prior. 

In Bayesian model selection, the evidence for a given model $\mathcal{M}$ under the Laplace approximation is the product of its goodness of fit and its Occam factor,
\begin{equation}
    \mathcal{P}(\mathbf{d}|\mathcal{M}) \ \simeq \ \mathcal{L}(\mathbf{d}|\theta_\text{MAP},\mathcal{M}) \ \pi(\theta_\text{MAP}|\mathcal{M})\sigma_{\theta|\mathbf{d}},
\end{equation}
where $\mathcal{L}(\mathbf{d}|\theta_\text{MAP},\mathcal{M})$ and $\pi (\theta_\text{MAP}|\mathcal{M})$ is the likelihood and prior evaluated at the best-fit of the model, and $\sigma_{\theta|\mathbf{d}}$ is the posterior uncertainty in parameters $\theta$. In essence, the Occam factor measures the information gained by the posterior from the collapse of the prior \citep{mackay2003information}, essentially punishing unnecessarily complex models for their superfluous predictive flexibility. This information gain is maximised under a prior which is considered to be maximally uninformative; priors exist which aim to satisfy this criterion, such as the reference prior \citep{bernardo1979reference,berger1992development,Berger_2009}, which reduces to the Jeffreys prior in one dimension. However, the implementation of a reference prior depends on how the parameters of interest are ordered and requires the determination of several conditional priors and the integration out of nuisance parameters, making it unfeasible in high-dimensional analyses. As we will see, the Occam penalty applied by a uniform prior is dictated by arbitrary parameter bounds, making model selection under this prior construct rather subjective. The Occam factor applied by the Jeffreys prior is discussed more in Section \ref{sec:results} when we compare the evidence obtained using a uniform prior and the Jeffreys prior.

As introduced above, the Jeffreys Prior is an improper prior. Bayesian model selection relies on the prior being proper, given that the Jeffreys prior is improper we make it proper by truncation by uniform bounds. The imposed bounds do introduce a level of subjectivity; therefore, the Jeffreys prior used in this work cannot be considered wholly objective; however, we argue that it is less subjective than the uniform prior due to the Jeffreys prior assigning prior weight based on Fisher information. A discussion on this particular aspect is given in Section \ref{sec:results}, but here we briefly note that expanding or contracting these bounds into regions of uninformative theoretical predictions has a diminishing effect (due to the small Fisher information) on the evidence integral. 

The practical implementation of the Jeffreys prior is computationally challenging, especially in high-dimensional spaces, as it relies on the determinant of the FIM and taking derivatives of a given model, this makes it susceptible to significant computation scaling. The implementation of such a prior is usually simplified by applying it only to nuisance parameters that appear linearly in a model, as done in EFTofLSS analyses \citep{donald2023analysis,gsponer2024cosmological}, in which the calculation then becomes trivial. Note that this is only considered a partial Jeffreys prior as a full Jeffreys prior will depend on both cosmological and nuisance parameters \citep{hadzhiyska2023cosmology}. The method of differentiation applied can play a consequential role in the performance of the Jeffreys prior. If one chooses finite differences, this requires determination of a suitable step size, which if too small suffers from numerical instabilities and if too large, a loss of accuracy. This method also requires several model computations per differentiation; therefore, it is not suitable for the computational overhead of implementing a Jeffreys prior. Instead, we choose to use automatic differentiation using \texttt{JAX} \citep{jax2018github}, which also benefits from Just-In-Time (JIT) compilation, allowing for extremely efficient computation and providing the main motivation for the development of a \texttt{JAX}-based pipeline in this work.

The calculation of the FIM is reduced from a second order derivative of the log-likelihood to a first order derivative of the model when we can assume Gaussianity of the likelihood. Under this condition, the FIM reads,
\begin{equation}\label{eq:GaussFish}
    \mathcal{F}_{ij}(\theta) = \frac{\partial \mu^T(\theta)}{\partial \theta_i}\Sigma^{-1}\frac{\partial \mu(\theta)}{\partial \theta_j}.
\end{equation}
Where \(\Sigma\) is the fixed data covariance matrix and \(\mu(\theta)\) is the theory vector corresponding to a particular dataset. We leverage this form of the FIM as we only consider likelihoods that are well approximated as Gaussian, hence we only require the derivative of the model with respect to the parameters. It is important to note that the dependence of the FIM on the data covariance makes the Jeffreys prior an experiment-dependent prior. We therefore interpret the Jeffreys prior as an alternative prior for the specific data combinations considered, rather than as a universal prior independent of the experiment.

\begin{figure}
    \centering
    \includegraphics[width=\linewidth]{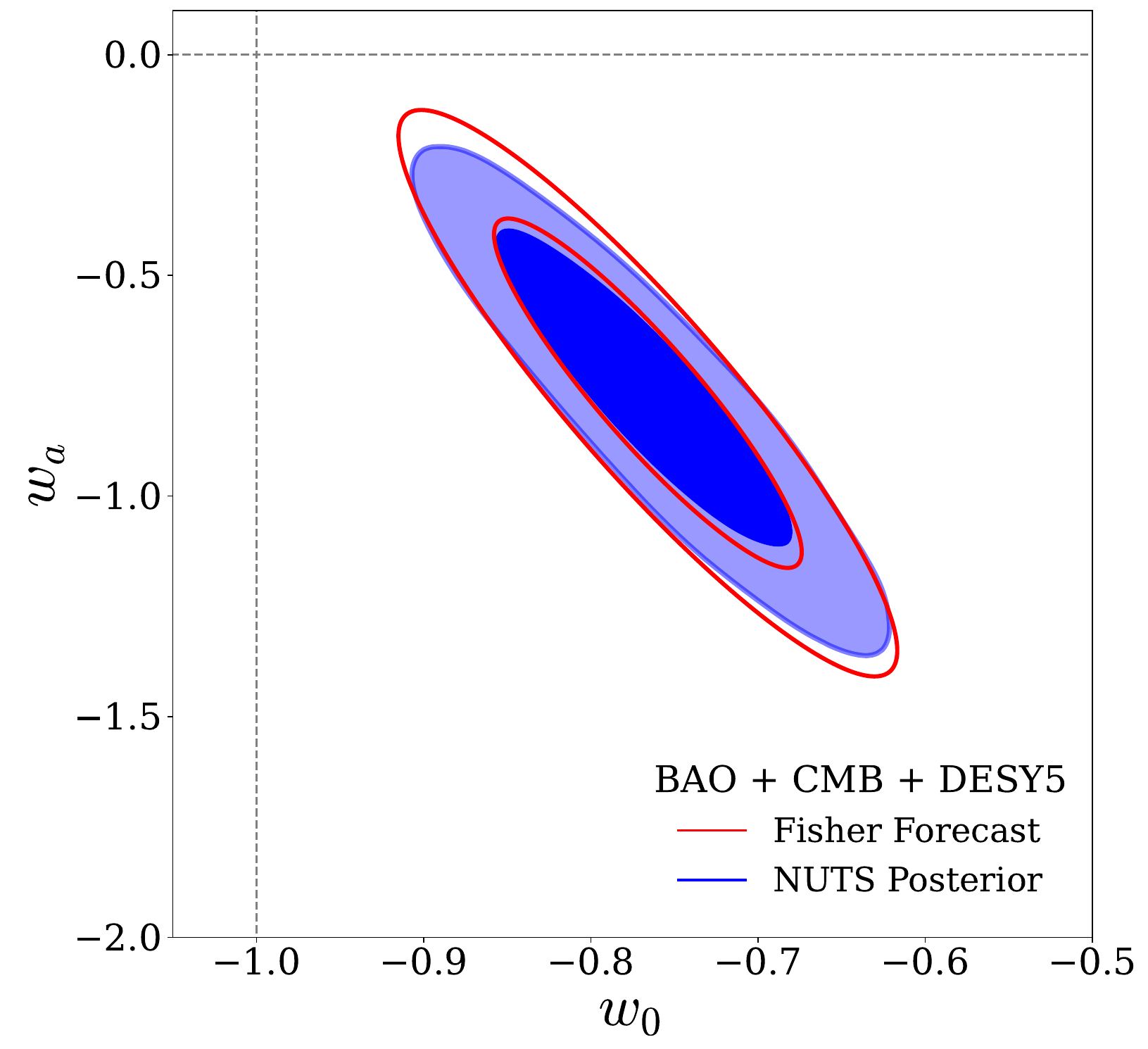}
    \caption{Fisher forecast versus the marginalised posterior contour obtained using the No-U-Turn Sampler (NUTS). The Fisher forecast is calculated using the Gaussian form of the Fisher information matrix (Equation \ref{eq:GaussFish}) and inverting this to use as the covariance matrix in a Gaussian distribution. The BAO data corresponds to DESI DR2 \citep{abdul2025desi}, while the CMB and SN data are the Planck 2018 high-\(\ell\) TTTEEE + low-\(\ell\) TT \& EE \citep{prince2022compressed} and DES-SN5YR \citep{abbott2024dark}, respectively. The consistent size and degeneracy direction between the two contours is sufficient evidence to suggest the Fisher information matrix is being calculated correctly.}
    \label{fig:FisherForecast}
\end{figure}
In Equation \ref{eq:GaussFish}, we evaluate the FIM under the assumption of a fixed data covariance matrix, \(\Sigma\), for each observable across the entire parameter space. Formally, if the covariance varies with the cosmological parameters, the FIM requires the inclusion of an additional trace term which depends on the derivatives of \(\Sigma\). Fortunately, neglecting the parameter dependence of the covariance matrix is a well-justified assumption in this analysis, as the information content is dominated by the signal derivative terms. This assumption holds across our independent datasets:
\begin{itemize}
    \item BAO - While both the shot noise and the shape of the underlying power spectrum dictate the variance, the fixed data covariance remains robust within the confines of our parameter space. In extreme regions of parameter space (\eg, \(\omega_b \rightarrow 0\)), the BAO signal would become virtually undetectable, which would severely inflate the true variance. However, our analysis is restricted to physically viable domains (\eg, \(\omega_b \in [0.015,0.03]\)); therefore, the BAO feature remains robust, the cosmological dependence of the covariance is minimal, and the information content is dominated by the signal derivatives.
    \item SN - The variance is driven by statistical noise and systematics rather than the underlying cosmology. Therefore, the dependence of the covariance matrix on the cosmological parameters will be weak, and the FIM will be dominated by the signal derivatives.
    \item CMB - Cosmic variance dominates the variance across most relevant scales and intrinsically scales with the theoretical power spectrum. However, as noted in early foundational CMB Fisher matrix literature (\eg, \citealt{Tegmark_1997}), the contribution of the parameter-dependent covariance trace to the total information content is sub-dominant compared to the sensitivity of the signal derivatives.
\end{itemize}

\begin{figure*}
    \centering
    \includegraphics[width=\linewidth]{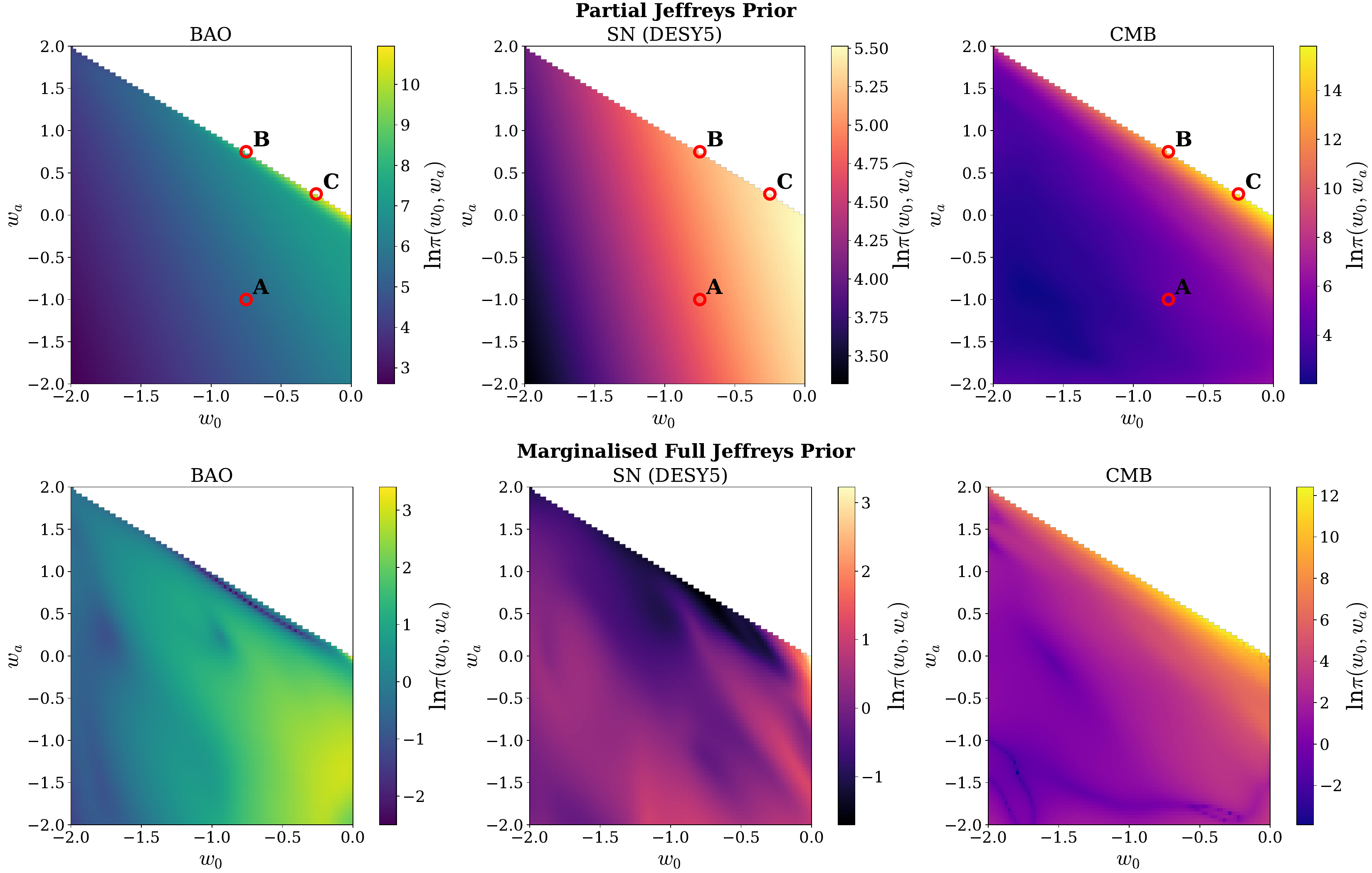}
    \caption{Top: partial Jeffreys prior computed with derivatives solely with respect to \(w_0\) and \(w_a\), and all other parameters set to fiducial \(\Lambda \)CDM values. Bottom: Jeffreys prior calculated by taking derivatives with respect to uniform prior parameters in the top of Table \ref{tab:priors} which are constrained by the respective dataset, and then analytically  marginalising over. BAO data corresponds to DESI DR2, while SN and CMB are DES-SN5YR and Planck 2018 TTTEEE \(+\) low-\(\ell\) TT \& EE, respectively. The structure observed in the marginalised prior is a consequence of marginalisation accounting for degeneracies between parameters. A hard prior boundary of $w_0+w_a<0$ is imposed to ensure early-time matter domination. The points A, B, and C represent the trajectory in \((w_0, w_a)\) space of Figure \ref{fig:JP_interp}.}
    \label{fig:JeffreysPrior_single}
\end{figure*}

The FIM calculation is straightforward to carry out for the respective datasets considered. For BAO this requires differentiation of the transverse comoving distance (\(D_M\)), the Hubble distance (\(D_H\)), and the volume averaged distance (\(D_V\)). SN consists only of the derivative of the distance modulus, and CMB of the angular power spectra. As the datasets are independent, the total FIM used in the Jeffreys prior is simply the sum of the individual data FIMs. 

In its Gaussian form the FIM becomes the inverse covariance matrix, which can be used as a sanity check to compare the Fisher forecast with the contour obtained from posterior samples. If the Fisher forecast aligns with the posterior it is satisfactory to deduce that the FIM is being calculated correctly. This test is shown in Figure \ref{fig:FisherForecast} for the dataset combination, BAO + CMB + DES-SN5YR. The degeneracy direction is consistent along with the size, suggesting that the FIM is realised accurately and the datasets used are well assumed to be Gaussian.

\subsection{Cosmological dependence of the Jeffreys prior for BAO, SN, and CMB data}
In Figure \ref{fig:JeffreysPrior_single} the Jeffreys prior on \(w_0\) and \(w_a\) is visualised for the different datasets. In order to show the Jeffreys prior and gain some insight from it, we calculate it using two alternative methods. First, we consider the partial Jeffreys prior (Figure \ref{fig:JeffreysPrior_single}; Top), which is computed by  differentiating solely with respect to \(w_0\) and \(w_a\) while keeping all other parameters at their fixed fiducial \(\Lambda\)CDM values. We find that the amplitude of the Jeffreys prior is large in regions of parameter space which force drastic changes in the expansion history of the Universe, with it generally peaking at $(w_0,w_a)=0$ for BAO, SN, and CMB. To understand the form of the Jeffreys prior in the ($w_0$, $w_a$) parameter space, first consider how the prior changes along a trajectory in this space starting at large negative values of ($w_0$, $w_a$) and moving towards the $w_0+w_a=0$ line. As we move towards this line, dark energy's early-Universe equation of state approaches that of matter (\(w\rightarrow 0\)). Consequently, the dark energy density scales similarly to matter, becoming a significant fraction of the early Universe's energy budget, which propagates through the observables as follows:
\begin{itemize}
    \item BAO - The presence of early dark energy significantly increases the Hubble parameter, \(H(z)\). This directly decreases the Hubble distance, \(D_H\), and suppresses the angular diameter distance, \(D_A\). Furthermore, the faster expansion rate means that sound waves have less time to travel before recombination, therefore shrinking the sound horizon at the drag epoch \(r_\text{drag}\). 
    \item SN - The significant increase in \(H(z)\) causes a corresponding decrease in the luminosity distance, \(D_L\), leading to high-redshift SN appearing brighter than in a \(\Lambda\)CDM cosmology.
    \item CMB - The angular scale of the acoustic peaks (\(\theta_\star\)) is heavily modified. Both the physical size of the sound horizon at recombination (\(r_\star\)) and the comoving angular diameter distance to the last scattering surface decrease. Because they shrink at different rates, the acoustic peaks shift. Additionally, the altered late-time expansion history modifies the decay of gravitational potentials, impacting the low-multipole power spectrum via the late Integrated Sachs-Wolfe (ISW) effect.
\end{itemize}

\begin{figure*}
    \centering
    \includegraphics[width=1\linewidth]{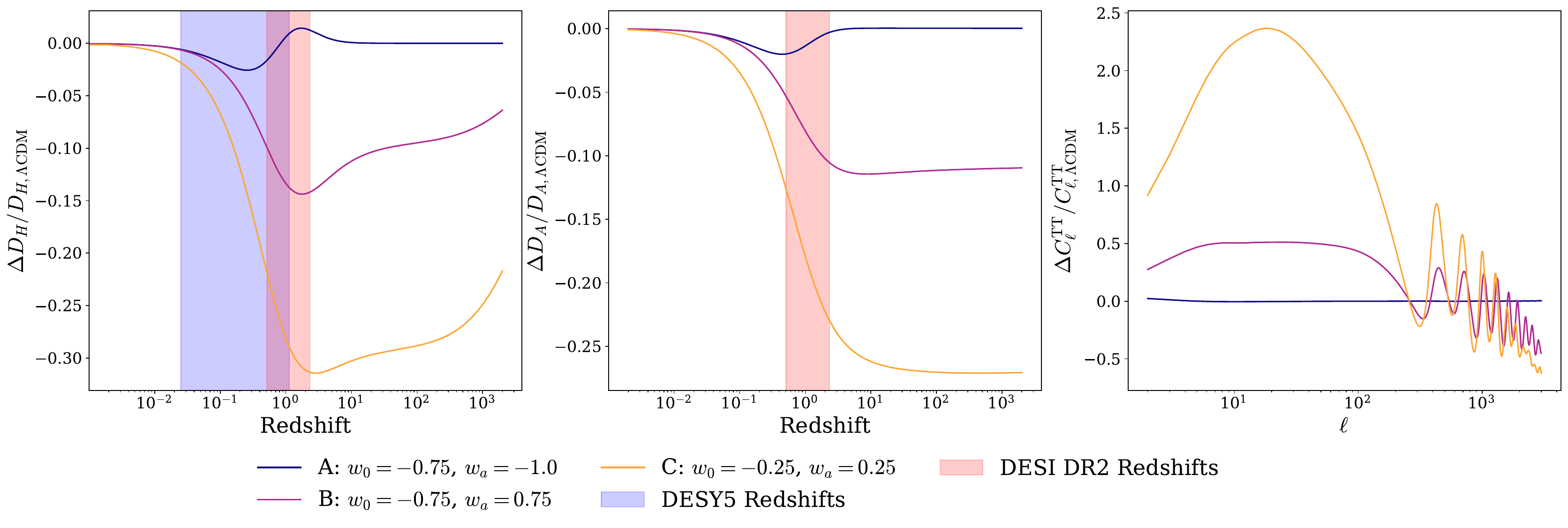}
    \caption{Changes to the observables used in this analysis when navigating the \((w_0,w_a)\) space. The specific trajectory in this space is shown by the corresponding points in Figure \ref{fig:JeffreysPrior_single}. The red and blue windows indicate the redshift ranges covered by the DESI DR2 BAO and DES-SN5YR SN data, respectively. The decrease in \(D_A\) and \(D_H\) is driven by the increase in the Hubble parameter due to an a higher prevalence of early dark energy as the \(w_0+w_a=0\) line is approached from large negative values in \((w_0,w_a)\) space. Moving along the \(w_0+w_a=0\) line towards \((w_0,w_a) = (0,0)\), dark energy becomes vastly more significantly in the early-Universe and late-time acceleration is weakened. The angular scale of the acoustic peaks in the CMB are heavily modified when dark energy becomes more significant in the early-Universe. Moving to less negative \(w_0\) causes changes to the decay of gravitational potentials; hence, having an impact on the low-multipole spectrum.}
    \label{fig:JP_interp}
\end{figure*}

Looking at these changes in the observables shown in Figure \ref{fig:JP_interp}, although the deviations in \(D_H\) and \(D_A\) as the $w_0+w_a=0$ line is approached (\(A\rightarrow B\) in Figures \ref{fig:JeffreysPrior_single} and \ref{fig:JP_interp}) are severe at high-redshift, they must naturally converge as \(z\rightarrow 0\), anchoring the models to local observations. Therefore, in the intermediate redshift ranges covered by the BAO and SN surveys, the theoretical distance curves must bend sharply to reconcile high-redshift deviations with the local anchor. To transition from the local Universe, where the model sensitivity to \(w_0\) and \(w_a\) is weak, to the early-Universe where the sensitivity increases due to increasingly diverging models, the derivatives at the intermediate redshifts must increase towards \(w_0+w_a = 0\), growing the Fisher information. These effects are further amplified when moving towards the \(w_0+w_a = 0\) line from regions of larger (less negative) \(w_0\), as the late-time acceleration is weakened.

Now, consider a second trajectory parallel to the $w_0+w_a=0$ line, moving towards the global maximum at $(w_0, w_a)=0$ (\(B\rightarrow C\) in Figures \ref{fig:JeffreysPrior_single} and \ref{fig:JP_interp}). This leads to a compounding of the above described effects. Not only does dark energy become vastly more significant at early times, but the late-time accelerated expansion is entirely dismantled as dark energy mimics matter across all cosmic epochs.

The bottom panel of Figure \ref{fig:JeffreysPrior_single} shows the Jeffreys Prior after analytically marginalising using a Gaussian approximation (\ie, using the Schur complement)\footnote{Figure \ref{fig:JeffreysPrior_single} presents two alternative forms of the Jeffreys prior; one where it is only applied to the \(w_0\)-\(w_a\) subspace of the parameter space, and one where it is applied to all parameters except calibration parameters, whose informative external priors are retained.}. We find that while the partial Jeffreys prior purely captures the sensitivity of the models to $w_0$ and $w_a$, it ignores the degeneracy between \(w_0\), \(w_a\), and other cosmological parameters. Instead through the act of marginalisation we observe the effect nuisance parameters and other cosmological parameters have on the Jeffreys prior; regions which previously had large information are now severely reduced by the nuisance parameters ability to mask changes. Conversely, the marginalised CMB prior retains its partial prior structure, demonstrating that the early-Universe deviations cannot be absorbed by the other cosmological and nuisance parameters.

\begin{figure*}
    \centering
    \includegraphics[width=\linewidth]{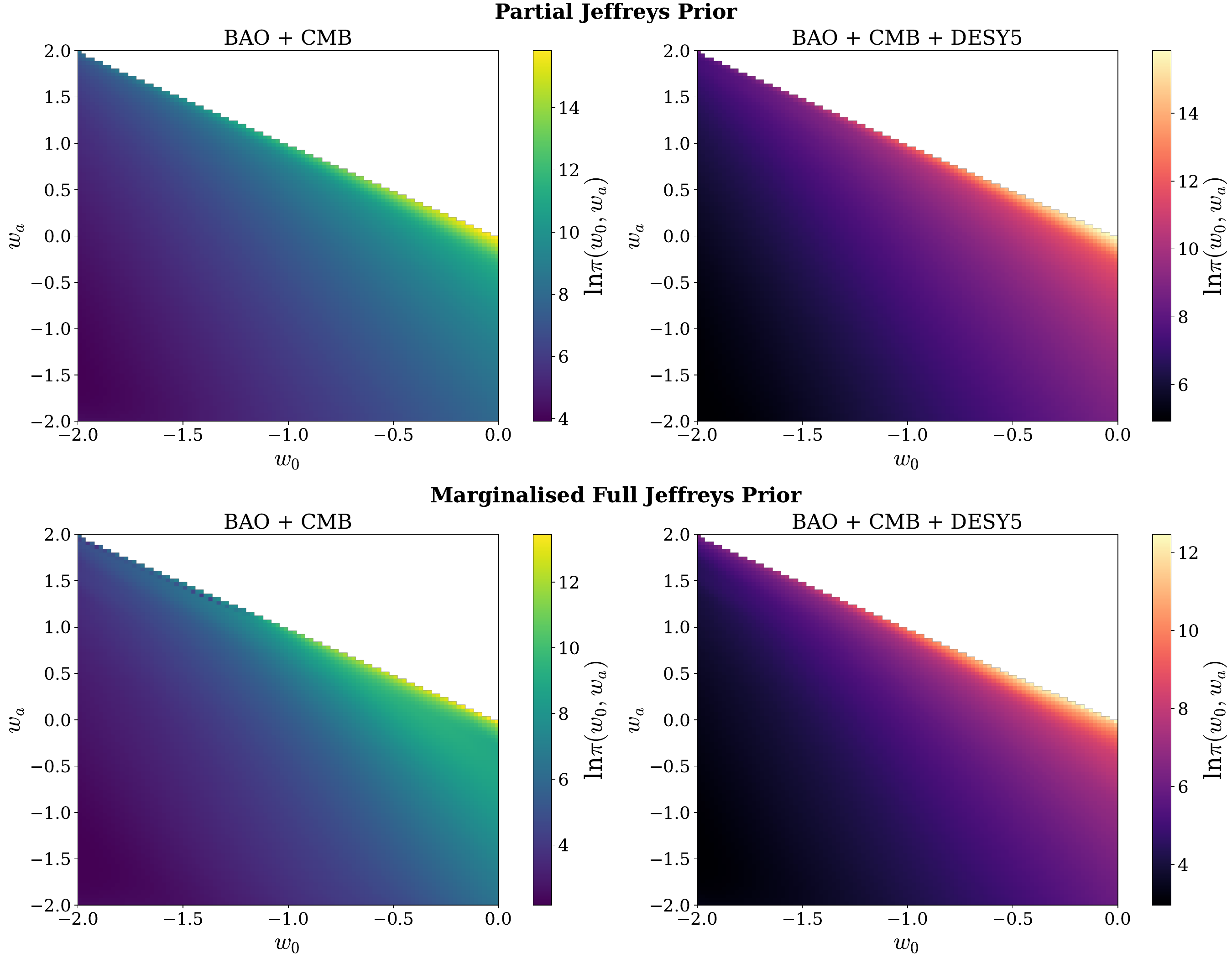}
    \caption{Top: partial Jeffreys prior computed with derivatives solely with respect to \(w_0\) and \(w_a\), and all other parameters set to fiducial \(\Lambda \)CDM values. Bottom: Jeffreys prior calculated by taking derivatives with respect to uniform prior parameters in the top of Table \ref{tab:priors} which are constrained by the respective dataset combinations, and then marginalising using the Schur complement. BAO data corresponds to DESI DR2, while SN and CMB are DES-SN5YR and Planck 2018 high-\(\ell\) TTTEEE \(+\) low-\(\ell\) TT \& EE, respectively. The diagonal line corresponds to the boundary \(w_0+w_a = 0\).}
    \label{fig:JeffreysPrior_combined}
\end{figure*}

When these two methods are applied to the dataset combinations we consider in this work (Figure \ref{fig:JeffreysPrior_combined}), the Jeffreys prior closely resembles the structure of the CMB prior. As the total FIM is the sum of the individual matrices, the larger Fisher information of the CMB exceeds the late-time probes. The tight constraints the CMB applies to nuisance parameters and other cosmological parameters prevents the late-time probes from using them to form degeneracy valleys, as previously observed (see Figure \ref{fig:JeffreysPrior_combined}), bringing the two methods closer to agreement.

\subsection{Cosmological Inference}\label{sec:CosmoInfer}
This work makes use of \texttt{JAX} for automatic differentiation, hence the gradient-based posterior sampler, the No-U-Turn Sampler (NUTS), from \texttt{numpyro} \citep{phan2019composable,bingham2019pyro} is used. This sampler relies on Hamiltonian dynamics to navigate the posterior, hence the name Hamiltonian Monte Carlo (HMC). Compared to the random walk Metropolis-Hastings algorithm used in \texttt{cobaya} \citep{torrado2021cobaya,lewis_cosmological_2002,lewis2013efficient}, NUTS is much more efficient at locating and sampling the target distribution. Its compatibility with \texttt{JAX} allows for a significant reduction in computation time, much needed in the application of a Jeffreys prior. The inclusion of gradients to explore the posterior can lead to divergences in complex distributions and at hard boundaries, to mitigate this the \texttt{target\textunderscore accept\textunderscore prob} is set around \(0.9\). NUTS is very efficient, so for most cases we run the sampler for \(2000\) warm up samples and a subsequent 4000 productions samples. This gives the sampler ample time to converge on the target distribution and sample from it. The sampling process is run with 4 chains in parallel with each chain assigned \(8\) CPUs. Chains are considered converged when the Gelman-Rubin criterion, \(\hat{R}-1<0.01\), is satisfied. This pipeline is considerably faster than \texttt{cobaya} at achieving convergence, with typical run times of \(\sim 30\) minutes for standard BAO + CMB + SN and \(\sim 3\) hours with a Jeffreys prior.

For theoretical predictions, we require a \texttt{JAX}-compatible package, so we use the cosmological emulator, \texttt{CosmoPower-JAX} \citep{piras2023cosmopower,SpurioMancini2022}. More information on how we trained the emulators is given in Section \ref{sec:Emulator}.

We keep the sampled parameters and their respective priors in accordance with the DESI Collaboration analysis as close as our setup allows. As the Jeffreys prior is improper, we cannot directly draw parameter proposals from it. To resolve this, we confine the parameter space using uniform prior distributions. The sampler draws proposal points from within these bounded uniform distributions, and the Jeffreys prior is evaluated at these specific points. The limits of these uniform distributions are dictated by the training bounds of our theoretical emulators (summarised in Table \ref{tab:priors}). Consequently, restricting the sampling to this bounded volume serves to truncate the Jeffreys prior, making it proper. The Jeffreys prior does not depend on calibration parameters which contain informative external priors. This includes the tight Gaussian prior given to the CMB power spectrum calibration parameter, \(A_\text{planck}\), motivated by the constraint given by \cite{aghanim2020planck}. The physical requirement, \(w_0 +w_a <0\), is enforced to allow for early-time matter domination, the same constraint applied in the DESI Collaboration analysis of DESI DR1 BAO data in \cite{adame2025desi}. We also investigate extensions to \(w_0w_a\)CDM including free curvature or free neutrino mass; our motivation being that these extensions open up the parameter space, which allows the prior to have an larger influence. The greater influence of the prior allows us to further test the robustness of dynamical dark energy under these different prior constructs. A summary of the parameter space and uniform bounds used is given in Table \ref{tab:priors}. 

\begin{table}
    \centering
    \begin{tabular}{lll}
    \toprule
        Parametrisation & Parameter & Prior/Bounds \\
        \midrule
        \bf{BAO\(+\)SN\(+\)CMB} & \(\omega_b\) & \(\mathcal{U}(0.015,0.03)\)\\
        & \(\omega_c\) & \(\mathcal{U}(0.09,0.15)\) \\
        & \(h\) & \(\mathcal{U}(0.4,1)\)\\
        & \(\ln(10^{10}A_s)\) & \(\mathcal{U}(2.5,3.5)\) \\
        &  \(n_s\) & \(\mathcal{U}(0.85,1.05)\) \\
        & \(\tau\) & \(\mathcal{U}(0.02,0.2)\) \\
        & \(w_0\) & \(\mathcal{U}(-2,0)\) \\
        & \(w_a\) & \(\mathcal{U}(-2,2)\) \\
        &  \(M_b\) & \(\mathcal{U}(-2,2)\) \\
        & \(A_\text{planck}\) & \(\mathcal{N}(1,0.0025)\) \\
        \midrule
        \bf{Extended} & \(\Omega_k\) & \(\mathcal{U}(-0.1, 0.1)\) \\
        & \(\sum m_\nu\) & \(\mathcal{U}(0, 0.5)\) \\
        \bottomrule
    \end{tabular}
    \caption{Sampled parameters and their respective priors. The prior bounds on cosmological parameters are based on the training range of the emulators used for theoretical predictions. \(M_b\) and \(A_\text{planck}\) are the SN and CMB calibration parameters respectively. The normal prior on \(A_\text{planck}\) is motivated by the constraints given in \cite{aghanim2020planck}. Parameters which have uniform priors are used in the Jeffreys prior, they define the truncation bounds of the Jeffreys prior and produce samples to be passed to the Jeffreys prior during inference.}
    \label{tab:priors}
\end{table}

To ensure that the pipeline used in this work reproduces results comparable to those of \texttt{cobaya}, the posterior distributions between the two methods are compared. The overlap between these contours is shown in Appendix \ref{sec:PipeVal}, in which they are in excellent agreement given the different pipelines and CMB data structure.

\subsection{Emulator Training} \label{sec:Emulator}
In order to make the most of the accelerated computation offered by \texttt{JAX} we use \texttt{CosmoPower-JAX} \citep{piras2023cosmopower, SpurioMancini2022} for theoretical predictions. We train two emulators, closely following the method used by \cite{jense2025complete}, to support either varying neutrino mass or curvature. 

We start by generating \(1.5 \times 10^5\) samples using Latin HyperCube (LHC) sampling to ensure an even distribution, with training bounds matching the uniform prior limits on cosmological parameters in Table \ref{tab:priors}. Theoretical predictions are generated from the samples using the Boltzmann solver \texttt{CAMB} \citep{lewis2000efficient,howlett2012cmb}, which are saved for emulator training. Table \ref{tab:EmTrain} summarises the geometric probes emulated and their respective ranges, which are set based on the limits of the BAO, SN, and CMB data. We note that some samples were rejected due to unphysical combinations of parameters, particularly  \(w_0 + w_a > 0\), which led \texttt{CAMB} to not be able to make predictions. The distribution of samples is checked to ensure reasonable coverage is still obtained in order for smooth training. The emulators are trained using \texttt{CosmoPower} \citep{SpurioMancini2022}, and later transformed to comply with \texttt{CosmoPower-JAX}. The temperature and polarisation power spectra are trained on their logarithms because they cover a large dynamic range, the logarithm compresses this range and creates a smoother landscape for the Neural Network (NN) to learn. The TE spectrum crosses zero, making the logarithm not a viable option, hence Principal Component Analysis (PCA) is used which acts as a linear compression. The settings we use when training match those used in \cite{jense2025complete}. The emulators we obtain are predominantly within \(1\%\) of \texttt{CAMB}, shown in the accuracy plots in Appendix \ref{sec:EmAcc}. The TE spectrum shows accuracy on a different scale due to its zero crossings and its values being very close to zero. The accuracy we achieve is deemed acceptable, especially with the agreement between our pipeline and \texttt{cobaya}.

Because we consider Bayesian evidence for model comparison, we wish to further reinforce our results by showing that changes to the evidence are not a consequence of emulator inaccuracy, therefore, we validate our emulators at the likelihood level compared to \texttt{CAMB}. We selected \(400\) random samples from the BAO + CMB + DES-SN5YR posterior and find the mean and maximum absolute difference between the two methods. For each sample, we calculate
\begin{equation}
    |\Delta \ln \mathcal{L}| = |\ln\mathcal{L}_\text{emulator} - \ln\mathcal{L}_\texttt{CAMB}|.
\end{equation}
In doing so we find a mean absolute difference of 0.07 and a maximum difference of 0.29 across the 400 samples. These differences are small in comparison to the changes in the Bayesian evidence induced by changing the prior, indicating that emulator inaccuracies are unlikely to be the cause of any changes observed in model preference.

\begin{table}
    \centering
    \begin{tabular}{lll}
    \toprule
       Quantity  & Range  & Emulator\\
       \midrule
       \(C_l^{TT}\)  & \(2 \leq l \leq 3000\) & NN of log-spectra\\
       \(C_l^{TE}\)  & \(2 \leq l \leq 3000\) & NN + PCA of spectra\\
       \(C_l^{EE}\) & \(2 \leq l \leq 3000\) & NN of log-spectra\\
       \(H(z)\)  & \(0 \leq z \leq 5\) & NN of evolution\\
       \(D_a(z)\) & \(0 \leq z \leq 5\) & NN of evolution\\
       \(r_\text{drag}\) & - & NN of value of \(r_d\)\\
       \bottomrule
    \end{tabular}
    \caption{Emulators trained using \texttt{CosmoPower-JAX} \citep{piras2023cosmopower,SpurioMancini2022}. The CMB TT and EE angular power spectra are trained on the logarithm of the spectra to reduce the dynamic range and create a smoother landscape for the Neural Network (NN) to learn. Principal Component Analysis (PCA) is used to compress the TE spectrum as the logarithm fails due to the zero crossings. The ranges chosen are to include the range of redshift and multipoles covered by the BAO, SN and CMB data. The training bounds coincide with the priors given in Table \ref{tab:priors}.}
    \label{tab:EmTrain}
\end{table}

\subsection{Model Comparison}
For a given model the Bayesian evidence is defined as 
\begin{equation}
    \mathcal{Z} = \int \mathcal{L}(\mathbf{d|\theta}) \pi(\theta) d\theta,
\end{equation}
where \(\mathcal{L}(\mathbf{d|\theta})\) is the likelihood and \(\pi(\theta)\) the prior. The log-evidence is related to the Kullback-Leibler divergence (KL) \citep{kullback1951information} as,
\begin{equation}
    \ln \mathcal{Z} = \langle\ln \mathcal{L}\rangle_\mathcal{P} - \mathcal{D}_\text{KL},
\end{equation}
where \(\langle\ln \mathcal{L}\rangle_\mathcal{P} = \int \mathcal{P}(\theta|\mathbf{d})\ln\mathcal{L}(\mathbf{d|\theta})d\theta\) is the posterior averaged log-likelihood. The above form explicitly highlights the functionality of the evidence as a measure of the balance between goodness-of-fit and the Occam penalty \(\mathcal{D}_\text{KL}\), where
\begin{equation}
    \mathcal{D}_\text{KL} = \int \mathcal{P}(\theta|\mathbf{d})\ln\frac{\mathcal{P}(\theta|\mathbf{d})}{\pi(\theta)}d\theta.
\end{equation}
The Occam penalty punishes models in which the collapse from prior accessible volume to posterior accessible volume is large, in other words models which are complex and require fine tuning to fit the data are more significantly penalised \citep{mackay2003information}. The Occam penalty relies not only on the dimensionality of the model but also on the prior probability assigned to parameters, thus making the evidence sensitive to prior choice. If one chooses to apply uniform priors, the prior volume is arbitrary as it relies only on the imposed prior bounds; hence, adopting wide flat priors inherently favours the less complex model. The Occam penalty applied by the Jeffreys prior is determined by how the Fisher information matrix weights the parameter space, and a more in depth analysis of the Occam penalty applied by the Jeffreys prior in this work is given in Section \ref{sec:results}. Neither prior construction is necessarily correct; however, the weighting of the parameter space using Fisher information lends itself to the Jeffreys prior being more objective than the uniform prior. 

The Bayes factor allows for the comparison between two competing models, in this work that is \(w_0w_a\)CDM and \(\Lambda\)CDM. The log Bayes factor is computed as
\begin{equation}
    \ln B = \ln\mathcal{Z}_{w_0w_a\text{CDM}} - \ln\mathcal{Z}_{\Lambda\text{CDM}}.
\end{equation}
A scale in which to interpret this result is the Kass and Raftery scale \citep{kass1995bayes} which is presented in Table \ref{tab:Kass}. 
\begin{table}
    \centering
    \begin{tabular}{ll}
    \toprule
       \(\ln B\)  & Evidence against \(\Lambda\)CDM\\
       \midrule
       \(0\) to \(1\)  & Not worth mentioning\\
       \(1\) to \(3\)  & Positive \\
       \(3\) to \(5\)  & Strong \\
       \(>5\)  & Very strong \\
       \bottomrule
    \end{tabular}
    \caption{Interpretation of log Bayes factor on the Kass and Raftery scale \citep{kass1995bayes} for the comparison between \(w_0w_a\)CDM and \(\Lambda\)CDM. Negative values suggest evidence in support of \(\Lambda\)CDM.}
    \label{tab:Kass}
\end{table}

The evidence is inherently expensive to compute. A common trick for nested models (such as $\Lambda$CDM, which is nested within $w_0w_a$CDM), is to use the Savage-Dickey Density Ratio (SDDR). However, we cannot use the SDDR, because the Jeffreys prior does not satisfy the conditional prior equality,
\begin{equation}
    \pi_{w_0w_a}(\phi|w_0=-1,w_a=0) \neq \pi_{\Lambda}(\phi),
\end{equation}
where $\phi$ are the common parameters of $\Lambda$CDM and \(w_0w_a\)CDM. Instead of the SDDR we use nested sampling, which requires the determination of the Jeffreys prior normalisation constant for each model. Without these constants, the final evidence ratio will differ from the true evidence ratio by the quotient between the normalising factors. The Jeffreys prior is generally improper; however, similar to posterior sampling, uniform priors must be used to sample from the parameter space, hence constricting the Jeffreys prior and making it proper. In order to determine the normalisation constant, we perform an initial nested sampling run with the unnormalised Jeffreys prior (\(\tilde{\pi}_\text{J}\)) as the likelihood. The same uniform priors (\(\pi_\text{U}\)) used in posterior sampling are used here (see Table \ref{tab:priors}) to create a bounded volume and sample from the parameter space. The evidence calculated by the nested sampler is then
\begin{equation}
    \mathcal{Z}' = \int_\Sigma \tilde{\pi}_\text{J}(\theta)\pi_\text{U}d\theta = \frac{1}{V_\text{U}}\int_\Sigma \tilde{\pi}_\text{J}(\theta)d\theta,
\end{equation}
where \(V_\text{U}\) is the enclosed volume defined by the uniform prior bounds and the subscript $\Sigma$ represents integration over the bounded subspace where $A_\text{Planck}$ is set to its fiducial value of $1.0$. The normalisation constant is then simply the evidence multiplied by this volume. Using this result in a secondary nested sampling run with the actual likelihood included gives the following expression for the evidence
\begin{equation}
    \mathcal{Z} = \int \mathcal{L}(\mathbf{d|\theta}) \frac{\tilde{\pi}_\text{J}(\theta)}{\mathcal{Z}'}\pi_\text{U}\pi_\text{N}d\theta = \int \mathcal{L}(\mathbf{d|\theta}) \pi_\text{J}(\theta)\pi_\text{N}d\theta,
\end{equation}
where \(\pi_\text{J}(\theta)\) is the normalised Jeffreys prior and $\pi_\text{N}$ is the normal prior on $A_\text{Planck}$. Hence, this setup provides the true evidence for a given model with a Jeffreys prior.

Applying this method allows for the determination of the evidence ratio between \(\Lambda\)CDM and \(w_0w_a\)CDM with a Jeffreys prior. To then compare to the uniform prior case, the same uniform bounds are used, and the Jeffreys prior is dropped, there is thus no need for an initial nested sampling run in the uniform prior case.

This method is carried out using the JAX-based nested sampler, \texttt{jaxns} \citep{albert2020jaxns}. A total of \(2000\) live points were used to provide a high resolution of the parameter space. We applied a termination condition to the sampler, in which if \(\ln(\mathcal{Z}_\text{current}-\mathcal{Z}_\text{remaining}) - \ln(\mathcal{Z}_\text{current}) < 0.01\) is satisfied, the sampling is ended and the resulting evidence is deemed sufficient. When sampling in the \(w_0w_a\)CDM parameter space, the physical boundary, \(w_0 + w_a < 0\), is set as in the posterior sampling to ensure early-time matter domination.

As \texttt{jaxns} saves the chains from the nested sampling run, these are used to calculate the posterior averaged log-likelihood which is then used in conjunction with evidence to calculate the Kullback-Leibler divergence. By comparing \(\mathcal{D_\text{KL}}\) between priors and models, we can deduce how significantly the dynamical dark energy claims are influenced by prior choice. 

\section{Results}\label{sec:results}
\subsection{Parameter Constraints}

\begin{figure}[h]
    \centering
    \includegraphics[width=\linewidth]{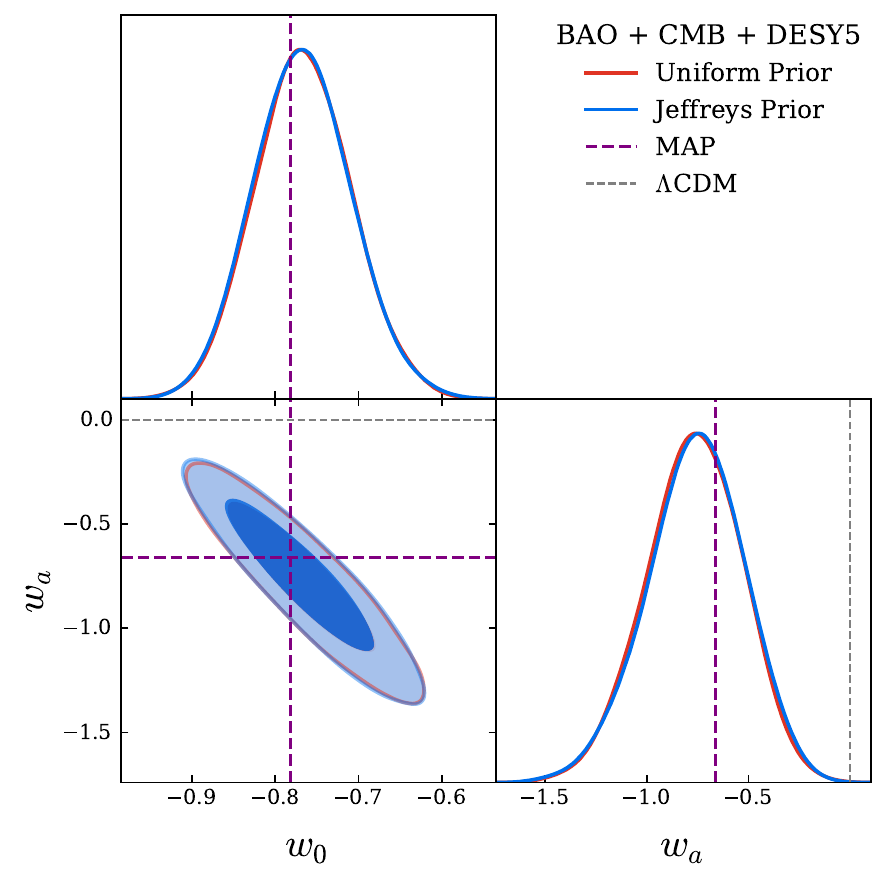}
    \caption{Plot of the \(w_0-w_a\) marginalised posterior for the combination DESI DR2 + Planck 2018 + DES-SN5YR. The uniform prior is shown in red while the Jeffreys prior is in blue. The purple dashed line is the \textit{Maximum A Posteriori} (MAP) computed using the uniform prior posterior. This is shown to be displaced from the marginalised posterior mode, indicating slight projection effects. The Jeffreys prior exhibits slight movement towards the MAP suggesting some action of lessening these effects.}
    \label{fig:uni_vs_jp(DESY5)}
\end{figure}

When the Jeffreys prior is applied to BAO + CMB + SN, the marginalised posterior is almost indistinguishable from the standard uniform prior case (see Figures \ref{fig:uni_vs_jp(DESY5)} and \ref{fig:uni_vs_jp(other)}). While we do observe offsets between the marginalised posterior mode and the \textit{maximum a posteriori} (MAP) estimate in the uniform prior case, which signifies some amount of projection effects occurring, these are not particularly significant. When considering the form of the Jeffreys prior in Figure \ref{fig:JeffreysPrior_combined}, the insensitivity of the marginalised posterior highlights the constraining power of the data itself. The largest movement appears with BAO + CMB, which can be attributed to the increased posterior volume as a result of less constraining power. Overall, the Jeffreys prior does appear to shift the marginalised posteriors towards the MAP, suggesting some mitigation of projection effects.

The ability of the Jeffrey's prior to lessen projection effects in the marginalised posteriors may be a direct consequence of the prior's structure shown in the bottom of Figure \ref{fig:JeffreysPrior_combined}. The resulting prior volume is almost indistinguishable from the stand-alone CMB prior (Figure \ref{fig:JeffreysPrior_single}; Right). As discussed in Section \ref{sec:jeffreysprior}, the Fisher information of the CMB dominates that from the late-time probes, preventing the formation of the degeneracy valleys we see in the bottom of Figure \ref{fig:JeffreysPrior_single}. This could be the reason for the minimal displacement of the marginalised posteriors towards the MAP, as the Jeffreys prior is primarily accounting for the projection effects present from CMB. This hypothesis is reinforced when we consider the BAO + CMB contours, as the projection effects from the CMB are more prevalent in this larger posterior volume.

In exploring extensions to \(w_0w_a\)CDM, we probe scenarios in which we increase the prior accessible volume and therefore the influence of the prior. We refer the reader to Appendix \ref{sec:extensions} for a more detailed discussion as we briefly summarise the results here. The same behaviour as in flat \(w_0w_a\)CDM is observed with free curvature, where no significant shift in the marginalised posteriors occur. The same narrative continues when neutrino mass is allowed to vary, where we also observe no movement in the neutrino mass posterior to more physical values.

In summary, the parameter constraints on \(w_0\) and \(w_a\) appear to be robust to adopting the Jeffreys prior, even when we open up the parameter space further or change dataset combination. The Jeffreys prior also shows little movement of the marginalised posterior mode towards the MAP estimate, however, this is not surprising as the Jeffreys prior is not designed to remove projection effects, it only tends to mitigate them.

\subsection{Model Selection}
The Bayesian evidence is known to be highly sensitive to the prior and bounds imposed on parameters. The application of the Jeffreys prior removes the arbitrary volume associated with uniform priors, instead weighting it by Fisher information. In this work, we computed the Bayes factor for BAO + CMB and BAO + CMB + SN, the results we obtain using the uniform prior and the Jeffreys prior are shown in Table \ref{tab:BayesFactor}. When the uniform prior is imposed, we find strong evidence for \(w_0w_a\)CDM over \(\Lambda\)CDM when considering DES-SN5YR. This substantially changes when we examine the recalibration, DES-Dovekie, in which the evidence does not favour either model. In the case of Union3, the preference for dynamical dark energy holds, although with weaker evidence, which is overturned by Pantheon+, showing positive evidence for \(\Lambda\)CDM. 

Our results are relatively consistent with previous research, such as \cite{ong2026bayesian} and \cite{hergt2026consistency}. We note that the prior bounds and the lack of CMB lensing cause our results to not be directly comparable to those of \cite{ong2026bayesian} and \cite{hergt2026consistency}. However, the results we acquired are widely in agreement with these previous works, and in the case of BAO + CMB, \cite{ong2026bayesian} considers no lensing and reports a value of \(\ln B = -0.07 \pm 0.28\), very similar to our value of \(\ln B = -0.06 \pm 0.18\). The agreement between our analyses, with each using their own respective pipelines, strengthens the validity of the results.

When we apply the Jeffreys prior, we find significant changes in the Bayes factor, demonstrating the sensitivity to prior choice. The strong evidence for dynamical dark energy suggested by DES-SN5YR, is entirely removed by the Jeffreys prior and replaced with very strong evidence in favour of the cosmological constant. A similar trend is observed throughout, where datasets supporting \(\Lambda\)CDM are further reinforced, while those that favour \(w_0w_a\)CDM now show a very strong preference for \(\Lambda\)CDM.

\begin{table}
    \centering
    \begin{tabular}{lcc}
    \toprule
        Datasets & \multicolumn{2}{c}{\(\ln B\)} \\
         & Uniform Prior & Jeffreys Prior \\
        \midrule
        BAO + CMB & \(-0.06\pm0.18\) & \(-11.54\pm0.18\)\\
        BAO + CMB + DESY5 & \(3.65\pm0.21\) & \(-5.52\pm0.21\) \\
        BAO + CMB + Dovekie & \(0.33\pm0.21\) & \(-9.18\pm0.21\) \\
        BAO + CMB + Union3 & \(2.34\pm0.19\) & \(-10.68\pm0.19\) \\
        BAO + CMB + PantheonPlus & \(-1.05\pm0.21\) & \(-10.57\pm0.21\) \\
        \bottomrule
    \end{tabular}
    \caption{Log Bayes factor for the given dataset combinations, a positive value indicates evidence in favour of \(w_0w_a\)CDM, while a negative value favours \(\Lambda\)CDM, where the amplitude is interpreted on the Kass and Raftery scale \cite{kass1995bayes} (see Table \ref{tab:Kass}). The given errors are propagated from the nested samplers statistical error. With CMB lensing neglected and different prior bounds, the uniform prior values are still relatively consistent with previous work \citep{ong2026bayesian,hergt2026consistency}. When the Jeffreys prior is applied, all combinations strongly support \(\Lambda\)CDM over \(w_0w_a\)CDM. The substantial change is driven by the more severe Occam penalty applied by the Jeffreys prior.}
    \label{tab:BayesFactor}
\end{table}

When we compute the Kullback-Leibler divergence (\(\mathcal{D}_\text{KL}\)) \citep{kullback1951information} (Table \ref{tab:stats}) we find that the significant swing in evidence in favour of \(\Lambda\)CDM are due to the Jeffreys prior applying a larger Occam penalty to the dynamical dark energy model. In the case of the uniform prior, the superior fit to the data is able to overcome the Occam penalty; however, this penalty depends directly on the subjective bounds imposed. As shown in \cite{patel2024prior}, decreasing the lower limit on \(w_0\) and \(w_a\) can reverse the preference for a particular model. The Jeffreys prior acts to avoid such an arbitrary assignment of punishment by weighting the prior accessible volume by the Fisher information, hence penalising highly distinguishable models where the posterior only resides within a very particular instance of predictions. Thus, our application of the Jeffreys prior demonstrates that the Bayesian preference for evolving dark energy is not robust to the choice of prior.

\begin{table*}
    \centering
    \begin{tabular}{llcccc}
    \toprule
         Datasets & &\multicolumn{2}{c}{\(\Lambda\)CDM} & \multicolumn{2}{c}{\(w_0w_a\)CDM} \\
        & &Uniform Prior & Jeffreys Prior & Uniform Prior & Jeffreys Prior \\
        \midrule
        BAO + CMB & \(\ln \mathcal{Z}\) & \(4945.38 \pm 0.12\) & \(4945.43 \pm 0.12\) & \(4945.32 \pm 0.13\) &  \( 4933.89 \pm 0.13\)\\
        & \(\mathcal{D}_\text{KL}\) & 21.04 & 21.00 & 23.74 & 35.11\\
        BAO + CMB + DESY5 & \(\ln \mathcal{Z}\) & \(5260.84 \pm 0.14\) & \(5260.13 \pm 0.14\) & \(5264.49 \pm 0.15\) &  \(5254.61 \pm 0.15\)\\
        & \(\mathcal{D}_\text{KL}\) & 26.14  & 26.84 & 30.38 & 40.04\\
        BAO + CMB + Dovekie & \(\ln \mathcal{Z}\) & \(5192.37 \pm 0.14\) & \(5191.64 \pm 0.14\) & \(5192.7 \pm 0.15\) &  \(5182.46 \pm 0.15\)\\
        & \(\mathcal{D}_\text{KL}\) & 26.12  & 26.91 & 30.05 & 40.07\\
        BAO + CMB + Union3 & \(\ln \mathcal{Z}\) & \(4982.66 \pm 0.13\) & \(4982.19 \pm 0.13\) & \(4985.00 \pm 0.14\) &  \(4971.51 \pm 0.14\)\\
        & \(\mathcal{D}_\text{KL}\) & 23.31  & 23.69 & 26.45 & 40.11\\
        BAO + CMB + PantheonPlus & \(\ln \mathcal{Z}\) & \(5783.76 \pm 0.14\) & \(5783.37 \pm 0.14\) & \(5782.71 \pm 0.15\) &  \(5772.80\pm 0.15\)\\
        & \(\mathcal{D}_\text{KL}\) & 26.21 & 26.63 & 30.11 & 40.00\\
        \bottomrule
    \end{tabular}
    \caption{Bayesian evidence (\(\ln \mathcal{Z}\)) and Kullback-Leibler divergence (\(\mathcal{D}_\text{KL}\)) \citep{kullback1951information} with uniform priors and the Jeffreys prior. The BAO data corresponds to the DESI DR2 data release, while the CMB data represents the Planck 2018 high-\(\ell\) TTTEEE and low-\(\ell\) TT \& EE likelihoods. The given errors on the log-evidence corresponds to the statistical error of the nested sampler.}
    \label{tab:stats}
\end{table*}

Our analysis does not establish that the Jeffreys prior is uniquely favoured, nor does it demonstrate that evolving dark energy is excluded by the data. Rather, we demonstrate that the model-selection conclusion obtained with the conventional uniform prior is reversed when considering an alternative prior, while posterior parameter constraints remain stable. Therefore, the evidence, in a strictly Bayesian interpretation, for dynamical dark energy should not be interpreted independently of the prior specification. We also emphasise that although the Jeffreys prior is invariant to alternative non-linear transformations to ($w_0$, $w_a$), our results still enforce a specific dark energy equation of state model which grows linearly with scale factor. Further work would therefore consider more flexible parametrisations for $w(a)$.

As the Jeffreys prior is improper, it must be made proper using truncation by uniform bounds. In order to test whether the strong preference for \(\Lambda\)CDM is an artefact of the truncation bounds, we perform evidence calculations using alternate bounds on \(w_0\) and \(w_a\) for the dataset combination BAO + CMB + DES-SN5YR. Due to the bounds already being at the limits of the emulator training, we evaluate the evidence across smaller sub-volumes. The first sub-volume considered is $w_0 \in [-1.5,0], w_a \in [-1.5,1.5]$, and the second is $w_0 \in [-1,0], w_a \in [-1,1]$. For the Jeffreys prior we find a difference between the fiducial bounds and first sub-volume of \(|\Delta \ln \mathcal{Z}| = 0.01\), while the evidence with the uniform prior differs by \(|\Delta \ln \mathcal{Z}| = 0.81\). Then in the second sub-volume we observe a \(|\Delta \ln \mathcal{Z}| = 0.1\) for the Jeffreys prior and \(|\Delta \ln \mathcal{Z}| = 1.36\) for the uniform prior. These results highlight the reduced sensitivity to prior bounds experienced by the Jeffreys prior, as the deviations are consistently within the nested sampling statistical error of \(\pm 0.15\). As the Jeffreys prior weights the enclosed parameter volume by Fisher information, the probability mass is heavily concentrated in regions of high theoretical sensitivity, expanding the prior bounds into geometrically large but theoretically uninformative regions has a diminishing effect on the evidence. As a consequence, the truncation bounds have minimal impact on the model preference under the Jeffreys prior. In contrast, the \(|\Delta \ln \mathcal{Z}| = 1.36\) experienced by the uniform prior is a significant difference considering the 3.65 log-Bayes factor given by the uniform prior. Other studies confirm such an impact, as \cite{patel2024prior} notes that larger bounds on \(w_0\) and \(w_a\) eventually result in a preference for \(\Lambda\)CDM. This further reinforces that the Bayes factor under the uniform prior is highly sensitive to the arbitrary bound choices, whereas the Jeffreys prior provides a substantially more stable measure of model preference. The Jeffreys prior forces \(w_0w_a\)CDM to account for its most extreme, distinguishable cases, and provides a semi-objective (given that it still relies on truncation bounds) and stable test of whether the model's superfluous predictive power is genuinely demanded by the data. 

\subsection{Pipeline and Prospects}
In implementing the Jeffreys prior developed a pipeline which is fully \texttt{JAX} compatible, employs binned CMB data, and uses gradient-based sampling, resulting in fast run times. An entire BAO + CMB + SN posterior inference run with standard uniform priors is able to converge within \(\sim 30\) minutes with only 32 CPUs, compared to \texttt{cobaya} which takes \(\sim 1\) day on 128 CPUs. Given this, improvements can still be made, the emulators used in this work could be trained to further accuracy. Training was predominantly limited by the computation time of \texttt{CAMB}, however, with the development of \texttt{JAX}-based Boltzmann solvers like \texttt{ABCMB} \citep{zhou2026abcmb}, emulator training can become significantly faster, or even replace the emulators entirely. 

The applicability of the Jeffreys prior in cosmology is contingent on the aims of the analysis. If one aims to effectively minimise projection effects, other methods exist, such as reparametrisation of nuisance parameters \citep{tsedrik2025simple,bonici2026alleviating}, however, this only remains effective if the nuisance parameters satisfy a linear relationship. The Jeffreys prior also seems most efficient at alleviating volume effects when it is applied only to nuisance parameters. Studies have used the Jeffreys prior on linear nuisance parameters in EFTofLSS  analyses \citep{gsponer2024cosmological,zhao2024multitracer,donald2023analysis,hadzhiyska2023cosmology} and shown it to mitigate projection effects. \cite{bonici2026alleviating} extended this to also apply the Jeffreys prior to non-linear nuisance parameters, where it exhibited better alleviation of projection effects than reparametrisation. Therefore, the Jeffreys prior on nuisance parameters remains a suitable choice in posterior inference, especially in the large nuisance parameter spaces of EFTofLSS analyses. However, the full Jeffreys prior implemented in this work does not seem applicable in this sense, as the added cosmological parameters introduce unnecessary added computation time with no added benefit of further mitigating projection effects. If the goal of the analysis is to remain objective the Jeffreys prior seems a suitable method to adopt, although it is computationally expensive. Alternative methods exist which avoid the multidimensional pathologies of the Jeffreys prior, such as the construction of reference priors which maximise the expected Kullback-Leibler divergence between prior and posterior \citep{trotta2017bayesian}. However, this approach still relies on the Fisher information matrix and the large computational overhead persists. An auto-differentiable implementation, such as that presented here, seems the only computationally feasible approach for the large parameter spaces characteristic of modern cosmological analyses.

\section{Conclusion}\label{sec:conclusion}
In this work, we investigated the impact of adopting an uninformative prior, the Jeffreys prior, on the evidence for dynamical dark energy. Instead of the conventional wide flat priors used in the DESI Collaboration analysis, we apply the Jeffreys prior to posterior inference and Bayesian evidence calculations.

Our first result is a null result; we found that when the Jeffreys prior is applied to posterior inference it has almost no effect on the constraints we measure for evolving dark energy. The largest shifts appear when the data is least constraining, opening up more of the parameter space, like we see for BAO + CMB. The lack of sensitivity to this particular change in prior strengthens the robustness of the posterior constraints on dynamical dark energy, such that, as we move into the Stage \RNum{4} era, the constraining power of data combinations like BAO + CMB + SN are becoming strong enough to overcome any major projection effects present in few-parameter extensions to $\Lambda$CDM.

Our work does highlight a stark contrast between posterior inference and Bayesian model comparison. Unlike the marginalised posteriors, the Bayesian evidence is highly sensitive to prior choice. When we move away from an arbitrary uniform prior towards a statistically motivated prior, the Jeffreys prior produces a consistent preference for \(\Lambda\)CDM across the dataset combinations considered. 

The Kullback-Leibler divergence revealed that the accordance between datasets was driven by the more severe Occam penalty imposed on \(w_0w_a\)CDM by the Jeffreys prior. The uniform prior applied an Occam penalty based on the collapse of the prior accessible volume to the posterior accessible volume; this evidently led to the dynamical dark energy model being penalised for the extra volume introduced by \(w_0\) and \(w_a\), which is entirely determined by the subjective bounds set. In the majority of cases we studied, this was not severe enough to overcome the dynamical models' superior fit to the data.  On the other hand, the Jeffreys prior was able to apply a statistically motivated Occam penalty, where the volume was weighted by the Fisher information. This led to the Jeffreys prior punishing \(w_0w_a\)CDM further for its distinguishable nature, where varying \(w_0\) and \(w_a\) produce large variations in model predictions of the expansion history, which can be strongly ruled out by BAO+CMB+SN data. As such, this collapse from a highly diverse parameter space to a posterior which inhabited only a small region meant the superfluous predictive power of \(w_0w_a\)CDM was not justified by the data.

We performed additional tests to establish the robustness of our results. Due to the Jeffreys prior being improper and must be made proper via truncation, we repeated evidence calculations using progressively tighter bounds on \(w_0\) and \(w_a\). The uniform prior evidence changes by up to \(|\Delta\ln\mathcal{Z}|=1.36\), whereas the Jeffreys prior evidence varies only by \(|\Delta\ln\mathcal{Z}|=0.1\), this demonstrates that the model preference obtained by the Jeffreys prior is substantially less sensitive to the arbitrary geometric truncation of the parameter space. We also tested the accuracy of our emulator at the likelihood level, by selecting 400 random posterior samples and finding the mean and maximum absolute difference between the emulator and \texttt{CAMB}. With a mean and maximum absolute difference of 0.07 and 0.29, respectively, there is no indication to suggest that the severe changes in the Bayes factor are the result of emulator inaccuracies and are instead solely induced by prior choice.

In summary, we have introduced novel techniques for interpreting the Jeffreys prior physically, which will be useful for future studies wishing to use this prior. Our results have demonstrated that the Bayesian evidence for evolving dark energy is not robust to prior choice. In particular, the preference obtained using the uniform prior is not reproduced when using the Jeffreys prior. We emphasise that this does not establish the Jeffreys prior as being uniquely favoured, nor does it rule out dynamical dark energy. Instead, it illustrates that claims of Bayesian evidence for dynamical dark energy should be interpreted together with prior specification.
\section*{Acknowledgements}
We thank Guadalupe Cañas Herrera and John Peacock for useful discussions. The authors acknowledge the support of a Royal Society University Research Fellowship. For the purpose of open access, the author has applied a Creative Commons Attribution (CC BY) licence to any Author Accepted Manuscript version arising from this submission.

\bibliographystyle{mnras}
\bibliography{refs}

\appendix

\section{Pipeline Validation}\label{sec:PipeVal}
To ensure the \texttt{JAX}-based pipeline we developed in this work is consistent with \texttt{cobaya}, we compare their marginalised posteriors. The marginalised posteriors shown below are in excellent agreement given the different MCMC method and CMB data structure.
\begin{figure}[h]
    \centering
    \includegraphics[width=0.5\linewidth]{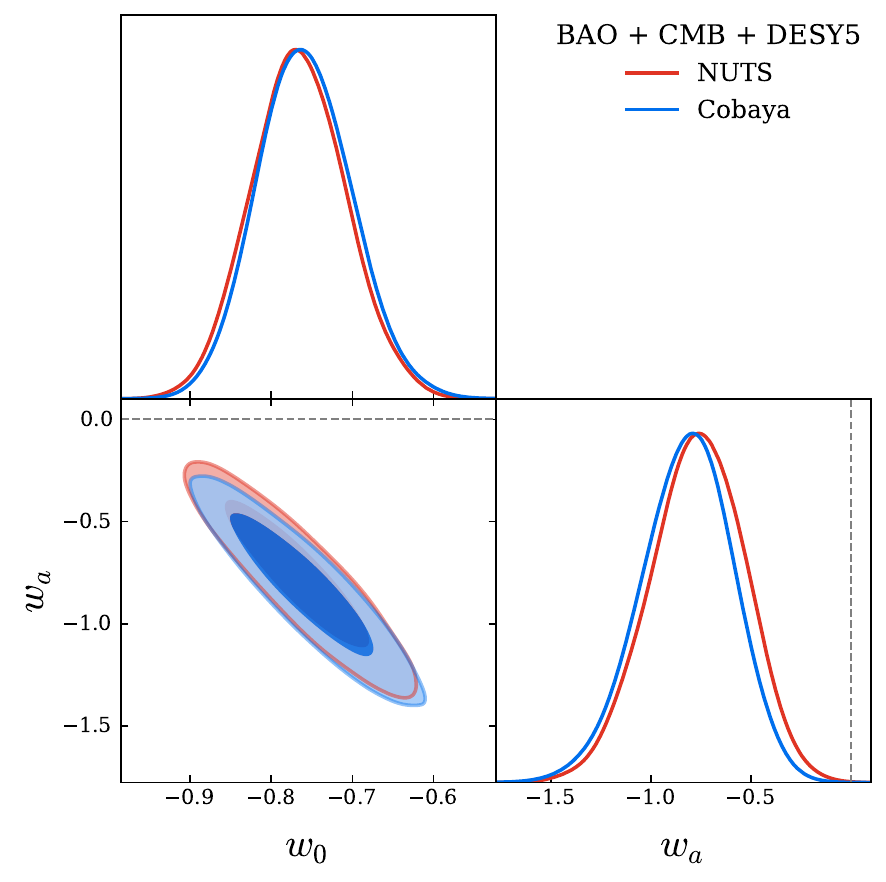}
    \caption{Comparison of the \(w_0\)-\(w_a\) contours between the \texttt{numpyros} NUTS sampler used in this work and \texttt{cobaya} used by DESI. Considering the different sampling method and Planck data implementation, the resulting contours are in good agreement. This is satisfactory to justify that the pipeline is in working order.}
    \label{fig:NUTSvCobaya}
\end{figure}

\section{Emulator Accuracy}\label{sec:EmAcc}
\begin{figure}[h]
    \centering
    \includegraphics[width=0.49\linewidth]{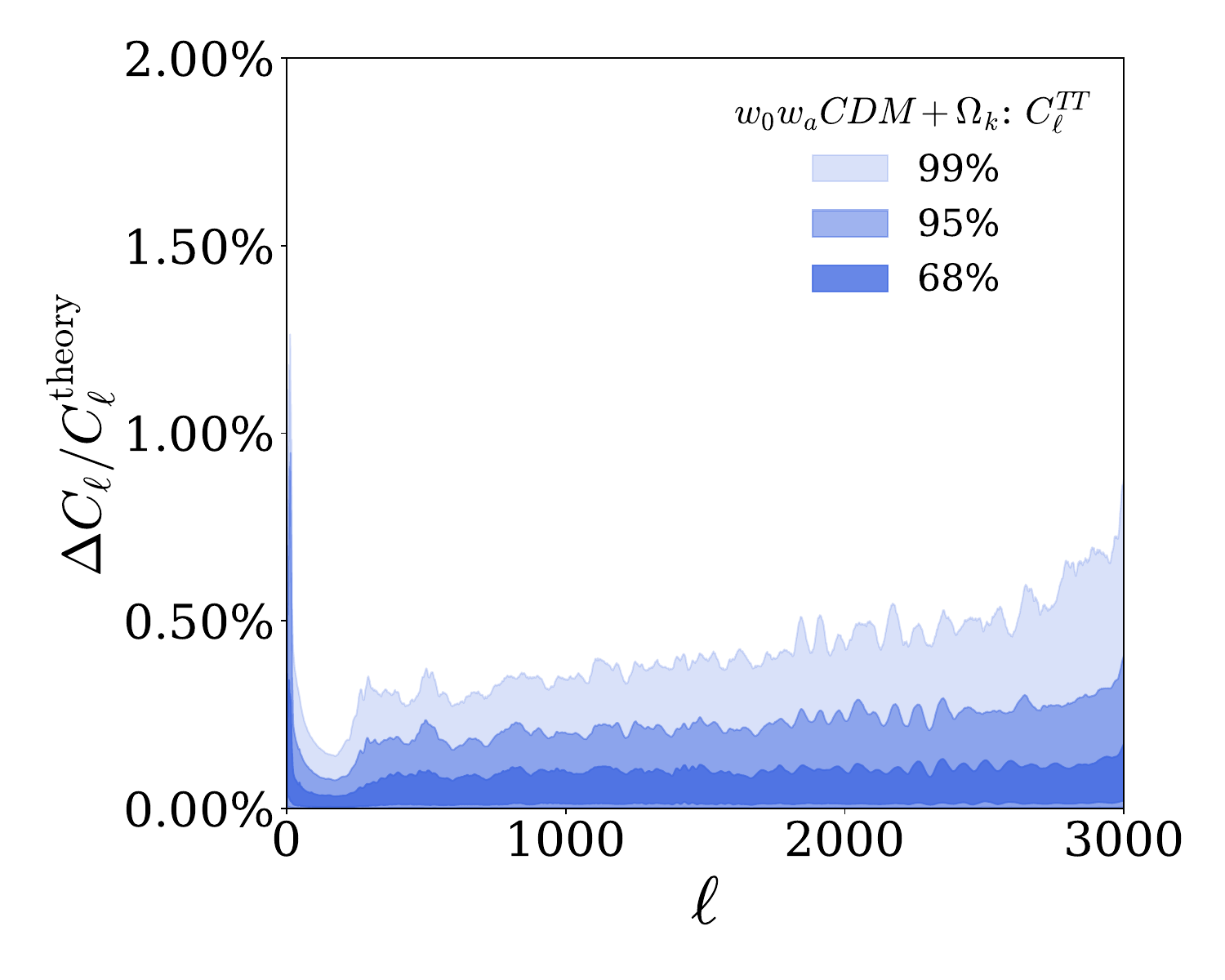}
    \includegraphics[width=0.49\linewidth]{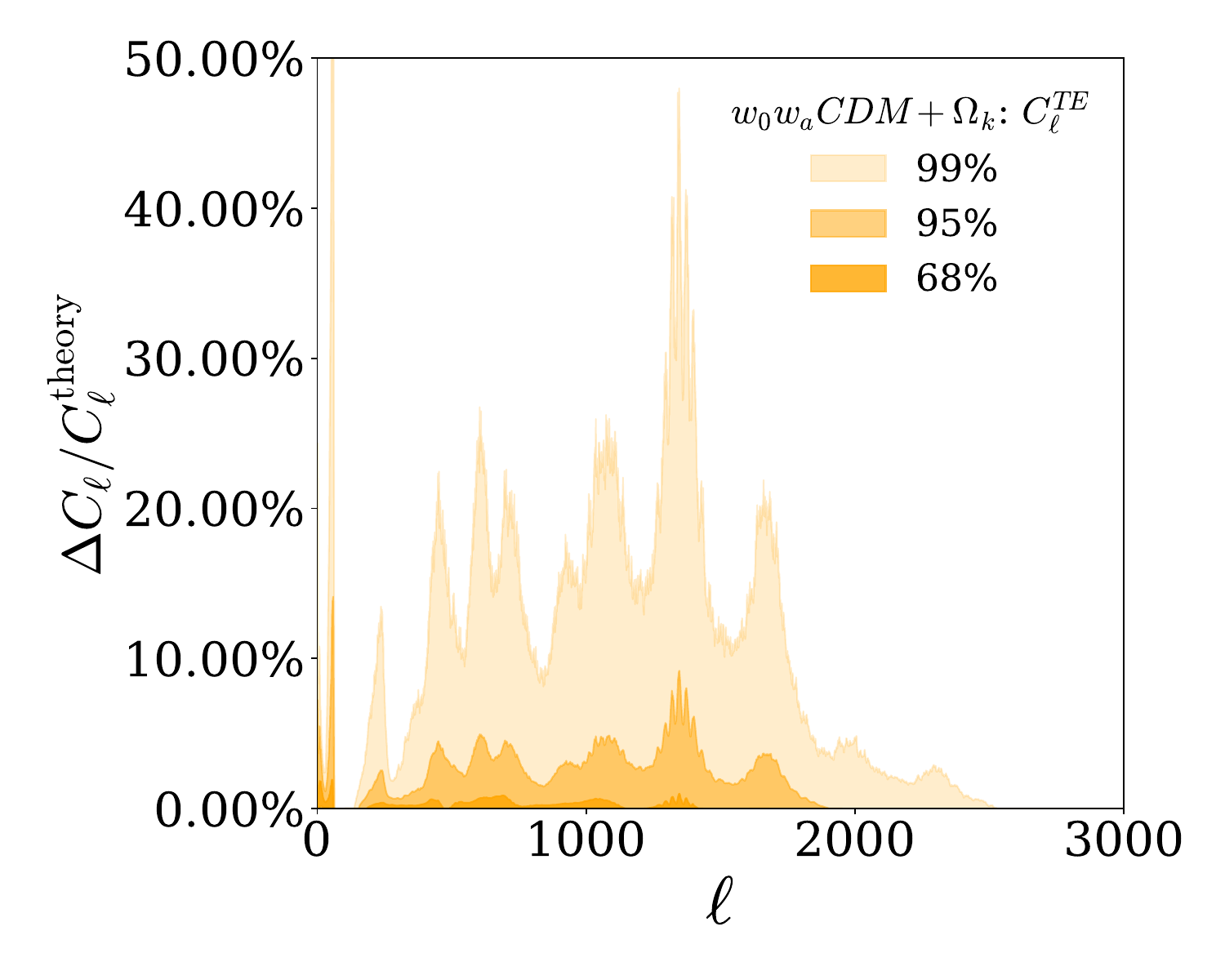}
    \includegraphics[width=0.49\linewidth]{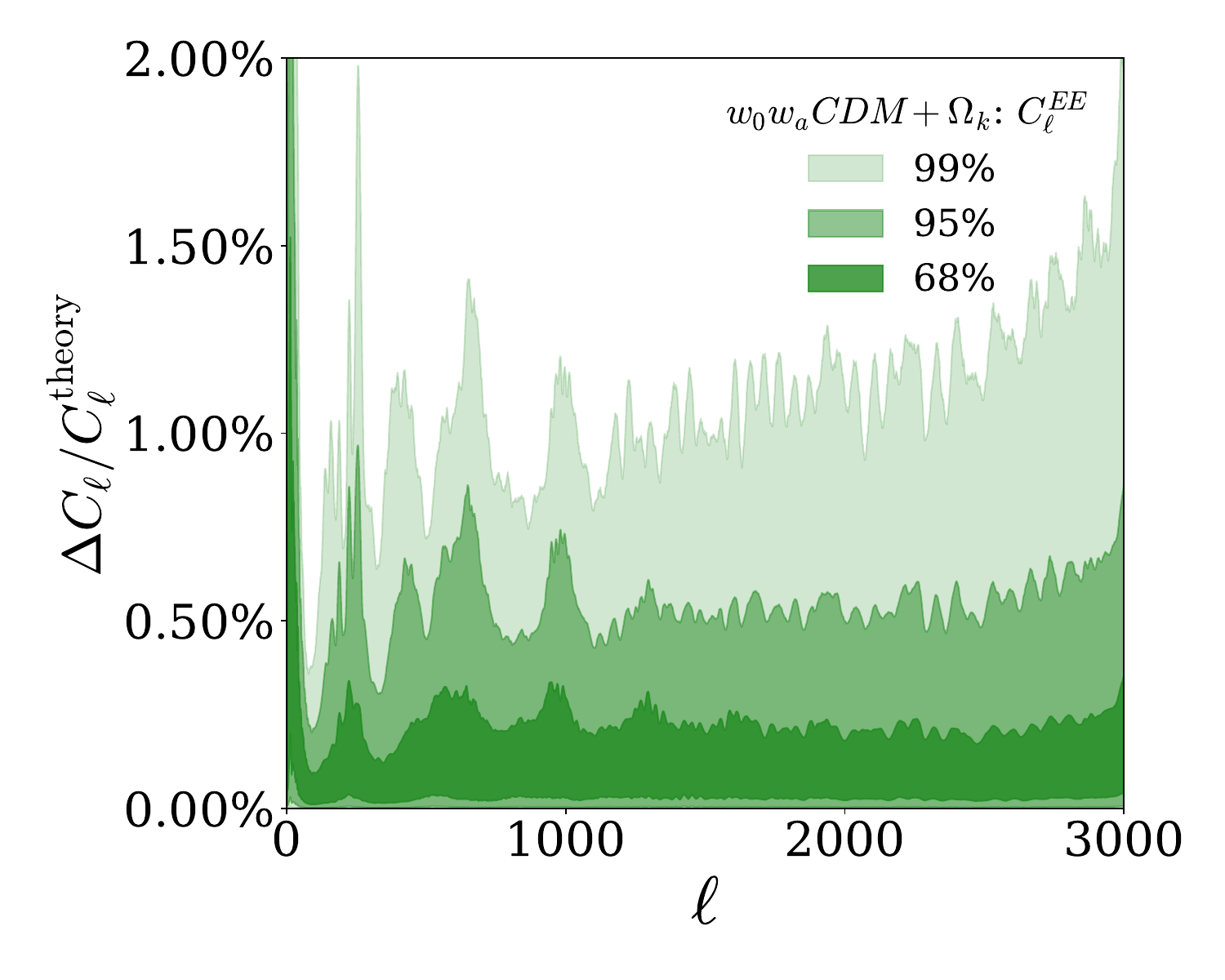}
    \includegraphics[width=0.49\linewidth]{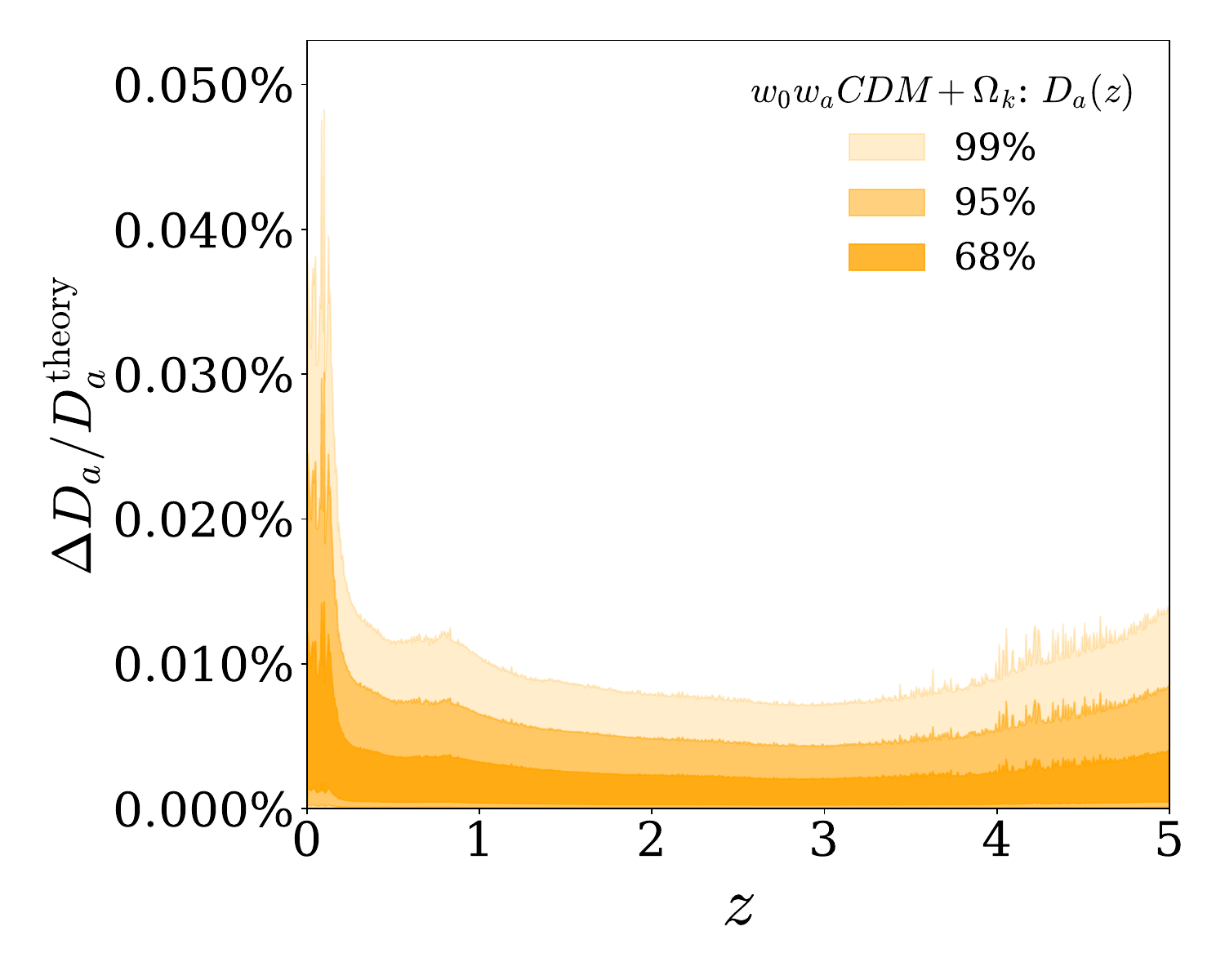}
    \includegraphics[width=0.49\linewidth]{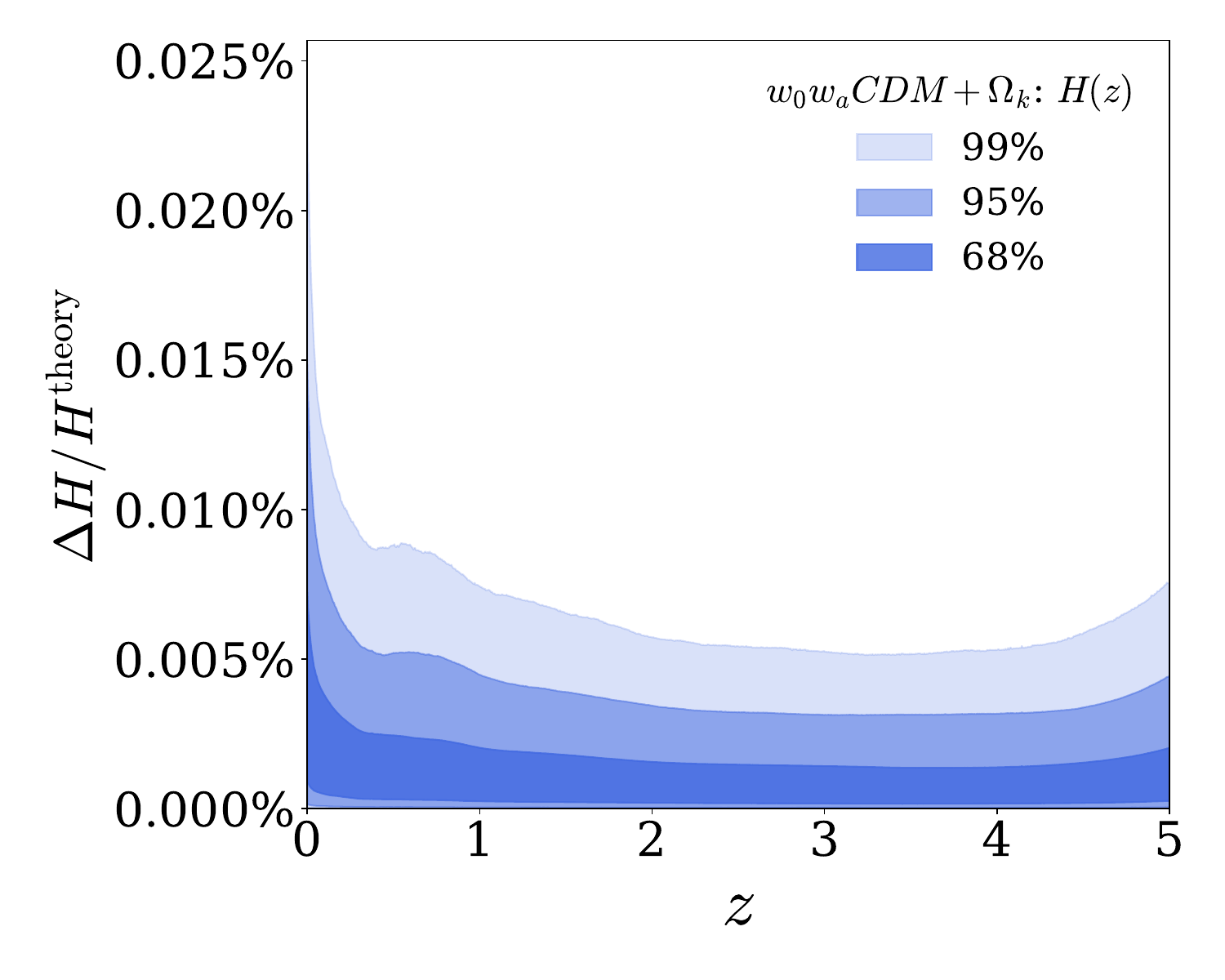}
    \caption{Plots of \texttt{Cosmopower-JAX} emulator accuracy compared to \texttt{CAMB}. The emulator was trained to allow for free curvature, with the training bounds corresponding to the uniform prior bounds on cosmological parameters given in Table \ref{tab:priors}. The training routine follows closely with that performed by \cite{jense2025complete}. Most emulators are within 1\% of \texttt{CAMB}, with the CMB TE emulator on a different scale due to the zero crossings.}
    \label{fig:Emacc1}
\end{figure}

\begin{figure}
    \centering
    \includegraphics[width=0.49\linewidth]{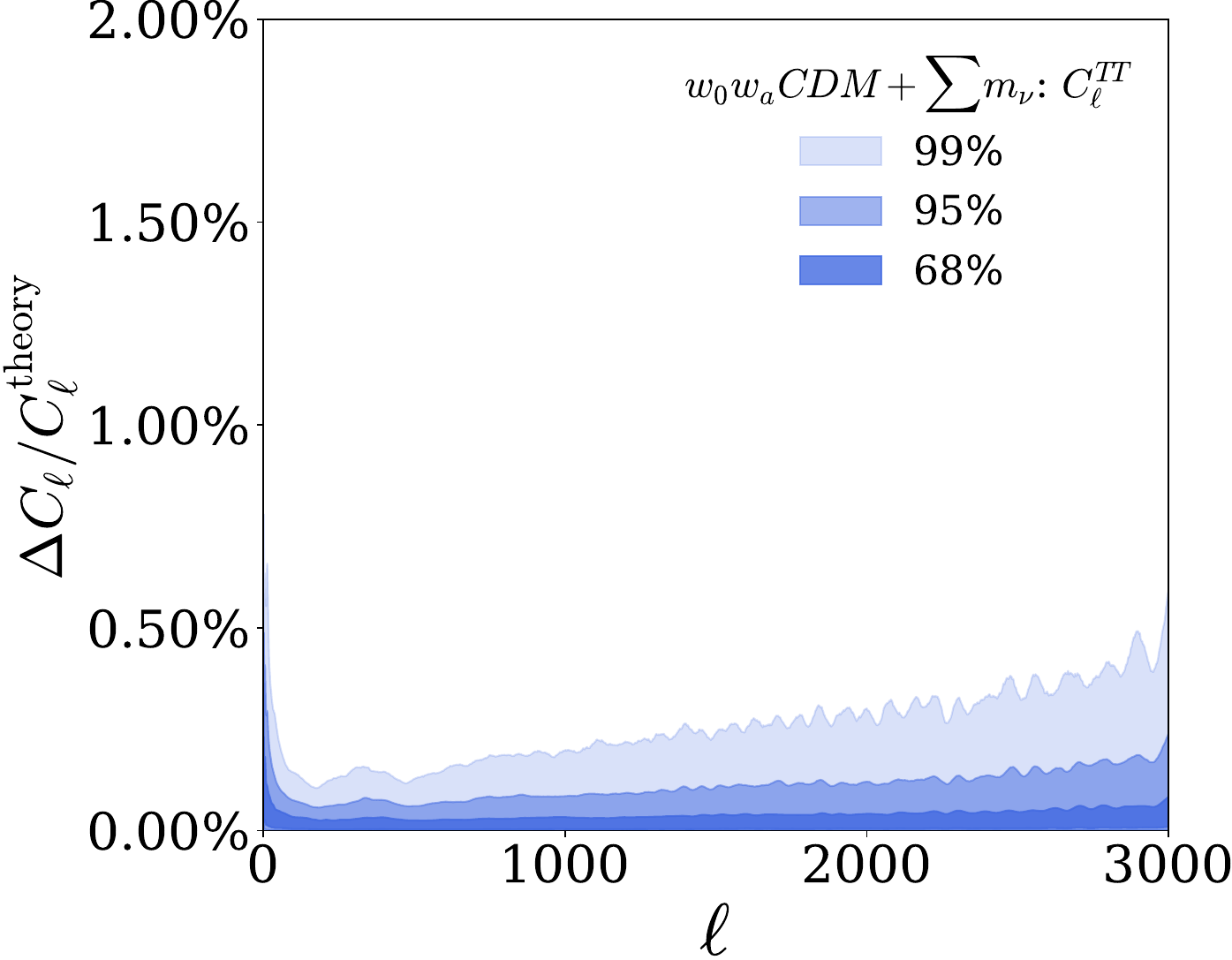}
    \includegraphics[width=0.49\linewidth]{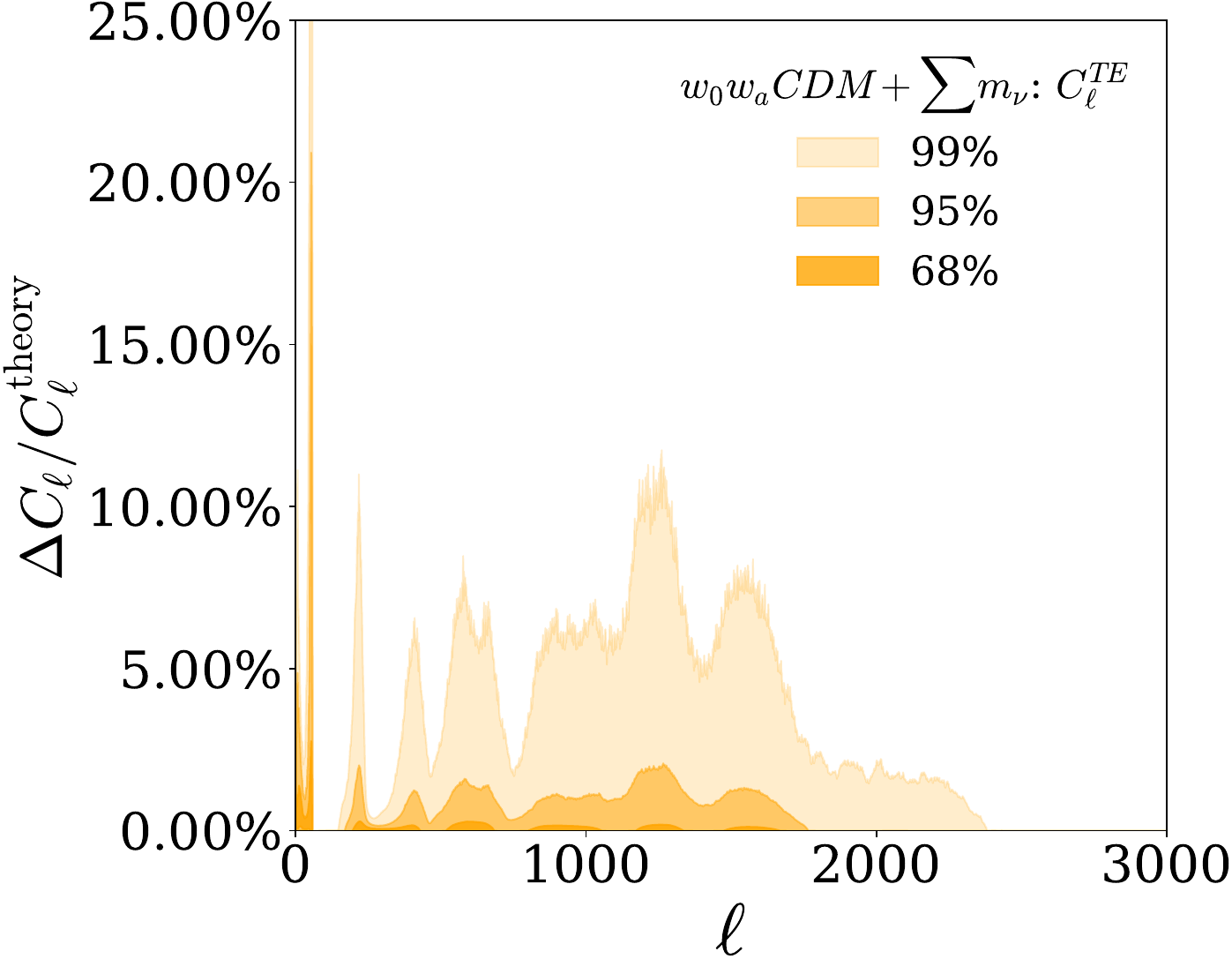}
    \includegraphics[width=0.49\linewidth]{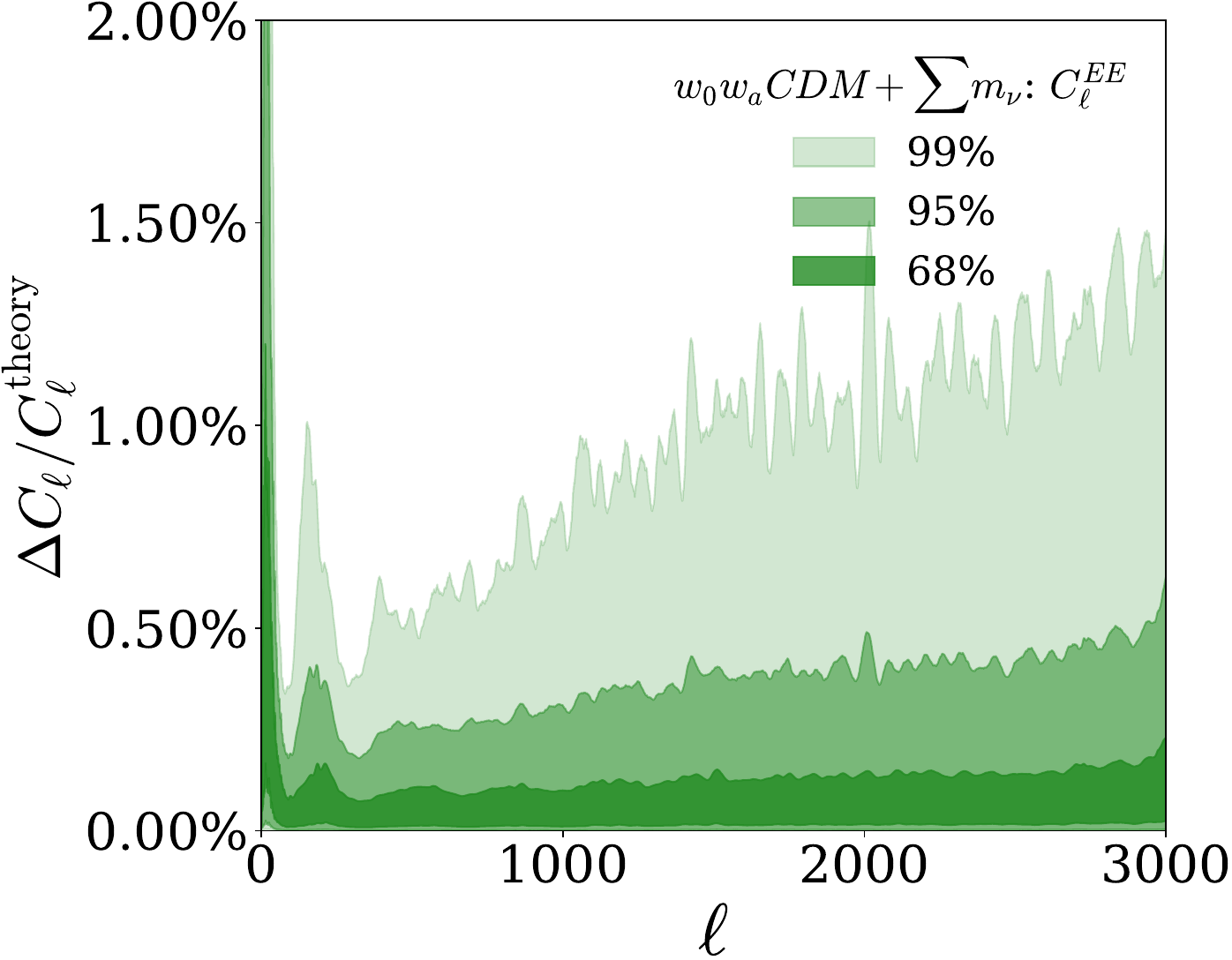}
    \includegraphics[width=0.49\linewidth]{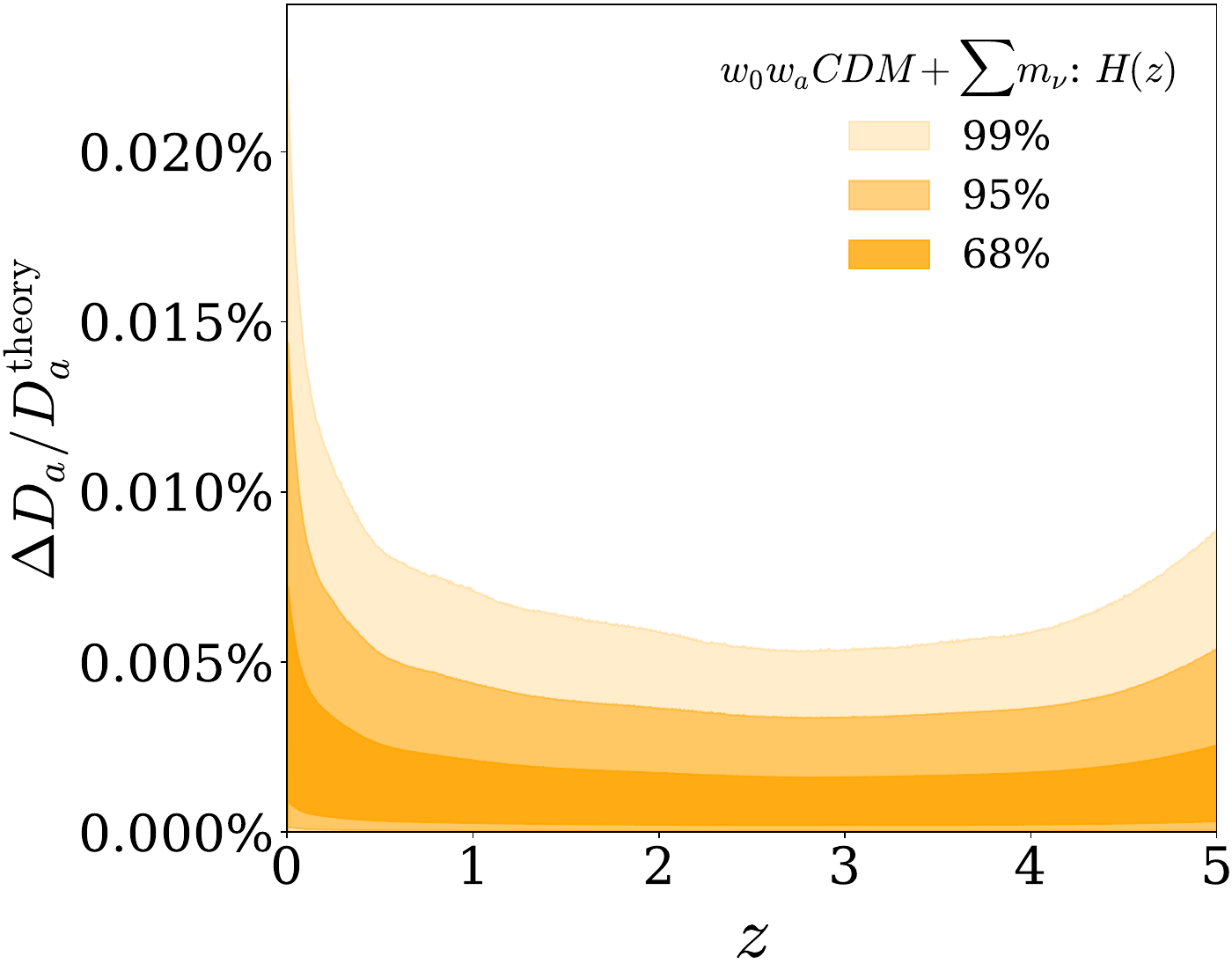}
    \includegraphics[width=0.49\linewidth]{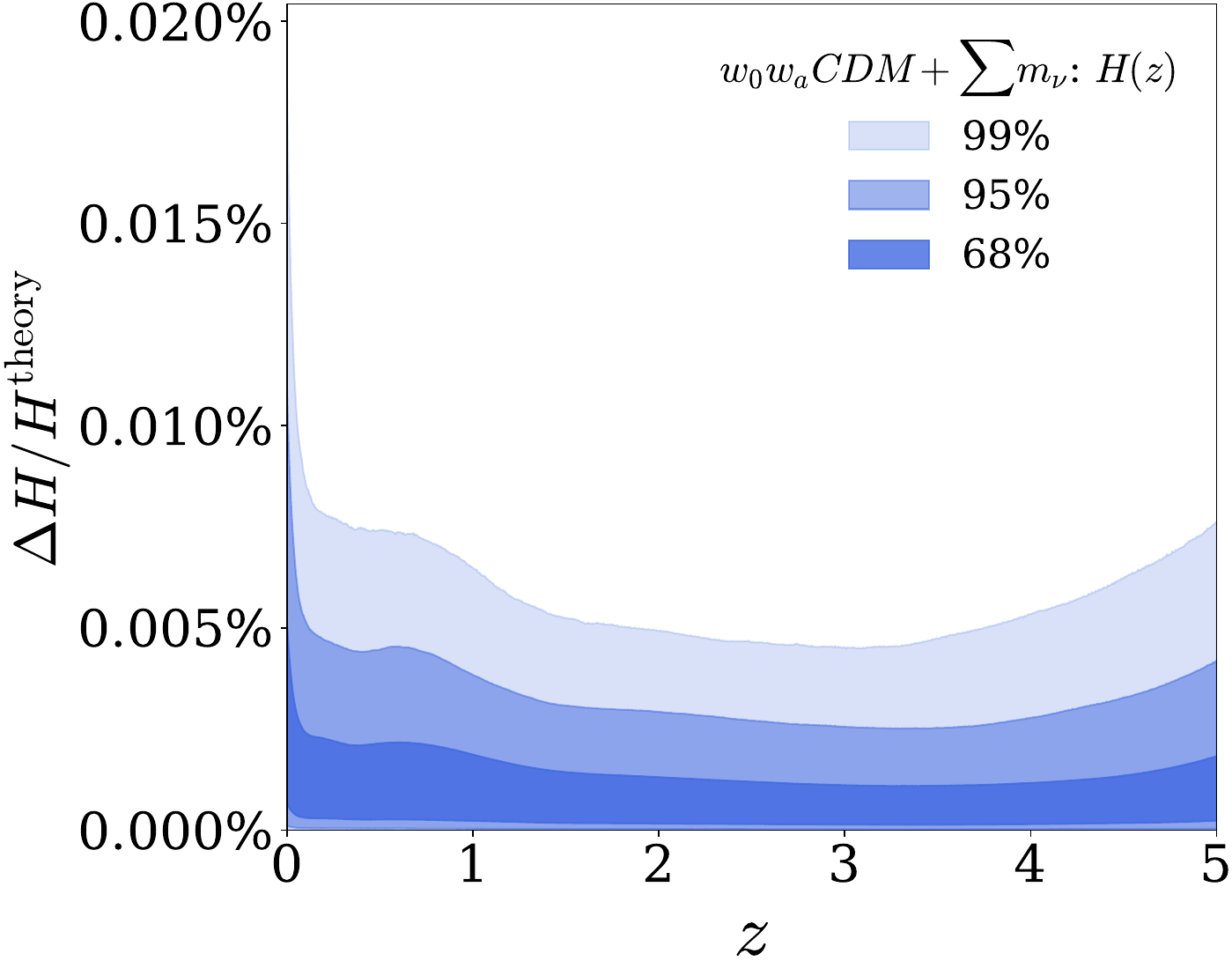}
    \caption{Plots of \texttt{Cosmopower-JAX} emulator accuracy compared to \texttt{CAMB}. The emulator was trained to allow for free neutrino mass, with the training bounds corresponding to the uniform prior bounds on cosmological parameters given in Table \ref{tab:priors}. The training routine follows closely with that performed by \cite{jense2025complete}. Most emulators are within 1\% of \texttt{CAMB}, with the CMB TE emulator on a different scale due to the zero crossings.}
    \label{fig:Emacc2}
\end{figure}

The fractional differences between the emulators used in this work and \texttt{CAMB} are shown in Figures \ref{fig:Emacc1} and \ref{fig:Emacc2}. Most of the emulators are within $1\%$ of \texttt{CAMB}, however, note that the inaccuracy in the CMB TE emulators are shown on a different scale due to the zero crossings. These emulators were trained using \texttt{CosmoPower} with the training routine closely following that outlined in \cite{jense2025complete}. The emulators were transformed to work with \texttt{CosmoPower-JAX}, which aided in the development of the \texttt{JAX}-based pipeline.

\FloatBarrier
\clearpage
\section{Alternative Dataset Combinations}\label{sec:alternateSN}
We also consider the effects of the Jeffreys prior on the posterior constraints of alternative SN data, such as Pantheon$+$, Union3, and the DES reanalysis DES-Dovekie. Similar to BAO + CMB + DES-SN5YR, we do not find any significant shift in the marginalised posteriors towards the \textit{maximum m posteriori} (MAP) estimate to suggest effective mitigation of projection effects. We also investigate the impact of the Jeffreys prior on BAO + CMB, where we observe the largest shift in the marginalised posteriors towards the MAP estimate. Note, that the marginalised posterior for \(w_a\) appears jagged; this is most likely a consequence of the No-U-Turn Sampler (NUTS) struggling with the hard boundaries of the parameters. The overall lack of movement in these additional posteriors under this alternate prior construct reinforces the reliability of the $w_0w_a$ constraints.
\begin{figure}[h]
    \centering
    \includegraphics[width=0.49\linewidth]{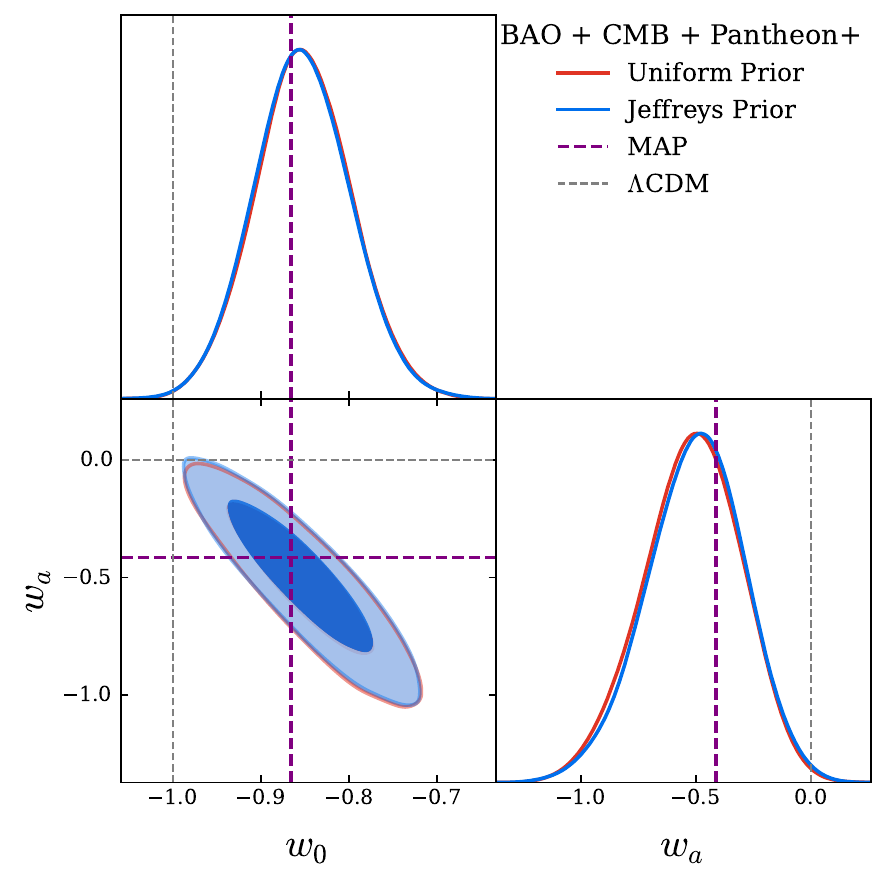}
    \includegraphics[width=0.49\linewidth]{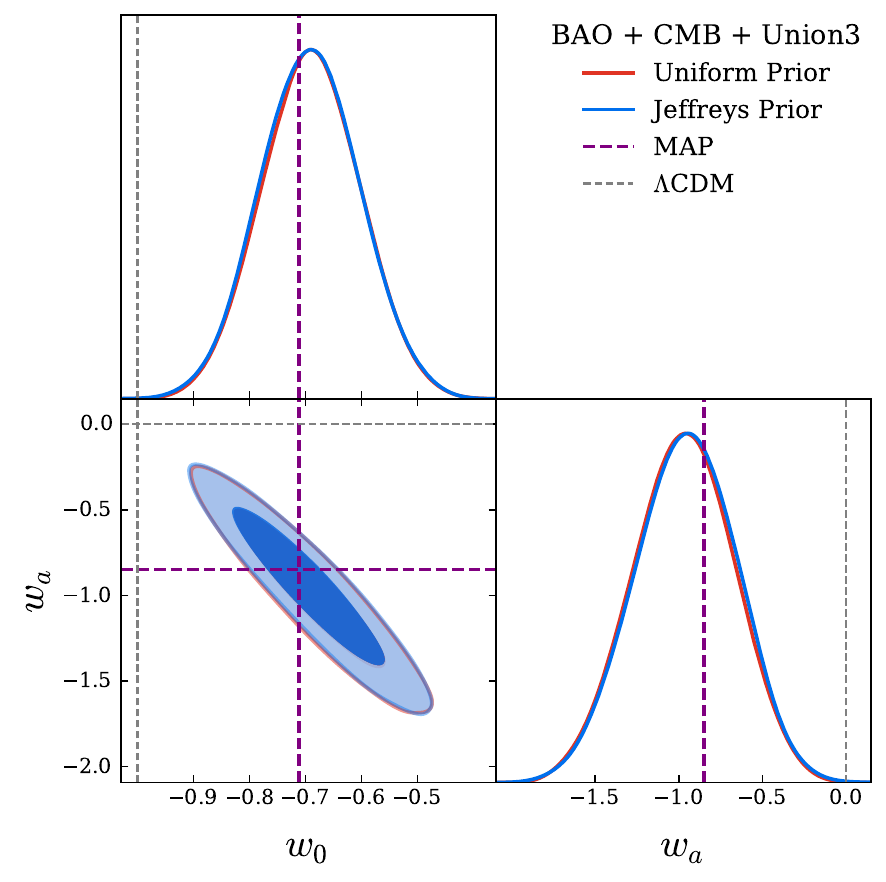}
    \includegraphics[width=0.49\linewidth]{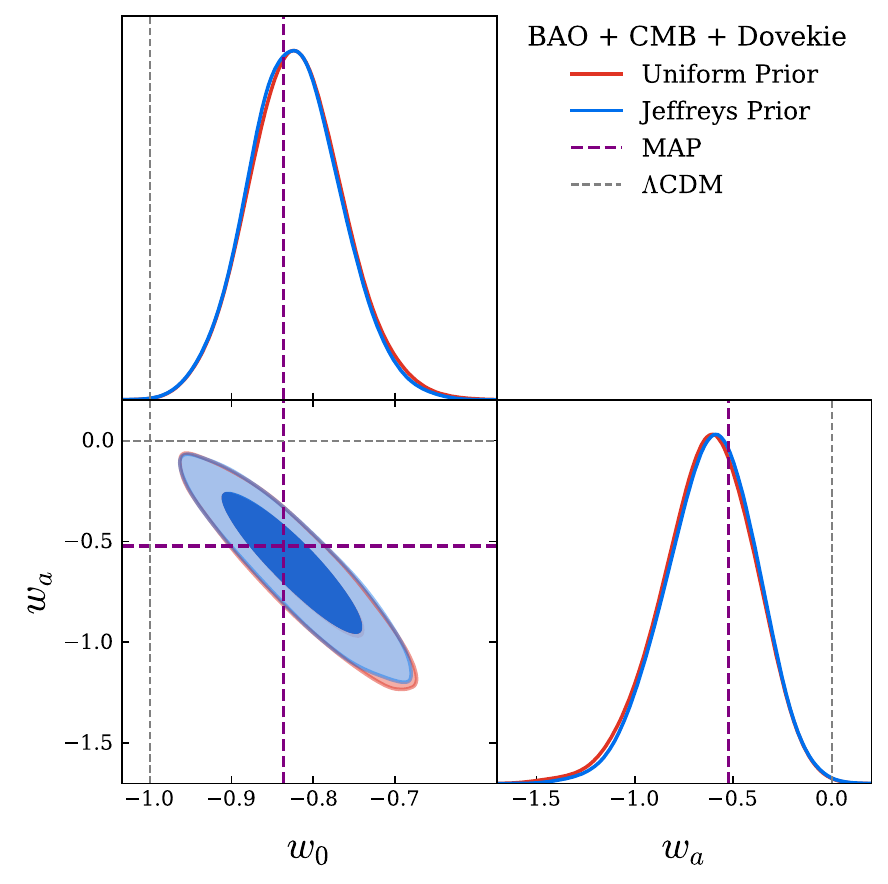}
    \includegraphics[width=0.49\linewidth]{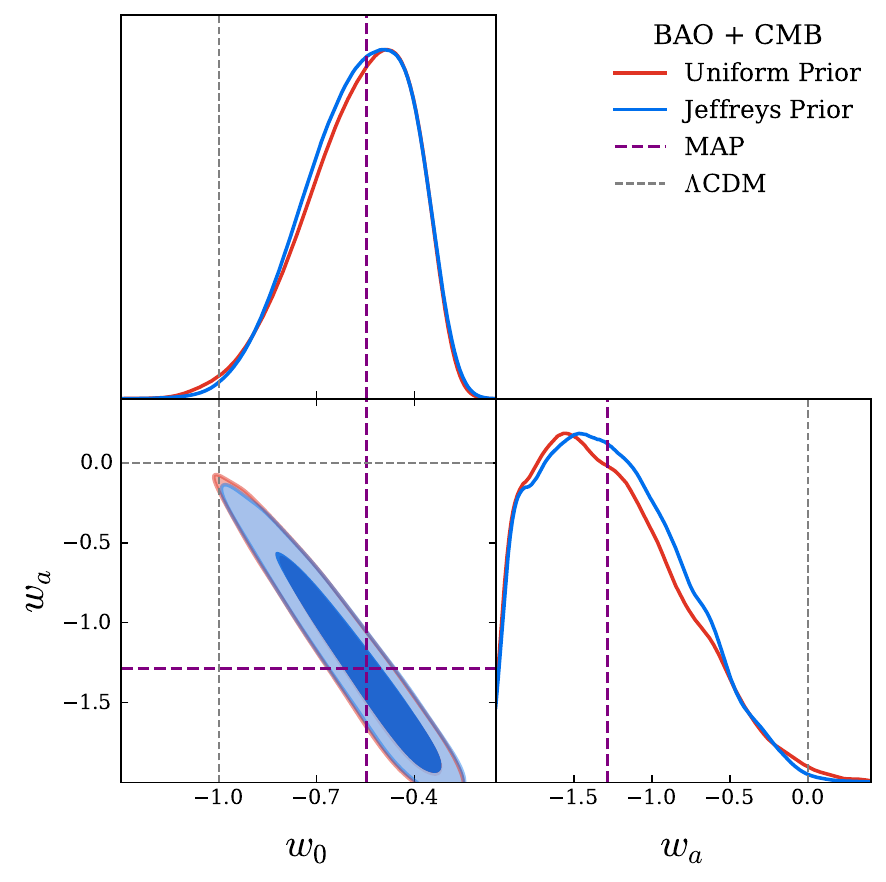}
    \caption{Plots of the \(w_0-w_a\) marginalised posterior with uniform priors (red) and a Jeffreys prior (blue) for varying SN datasets and the omission of SN data. The purple dashed line is the \textit{Maximum A Posteriori} (MAP) estimate computed using the uniform prior posterior. This is shown to be displaced from the marginalised posterior mode, indicating slight projection effects. The Jeffreys prior exhibits slight movement towards the MAP suggesting some action of lessening these effects.}
    \label{fig:uni_vs_jp(other)}
\end{figure}
\FloatBarrier
\newpage

\section{Extensions to \(w_0w_a\)CDM}\label{sec:extensions}
\begin{figure}[h]
    \centering
    \includegraphics[width=0.495\linewidth]{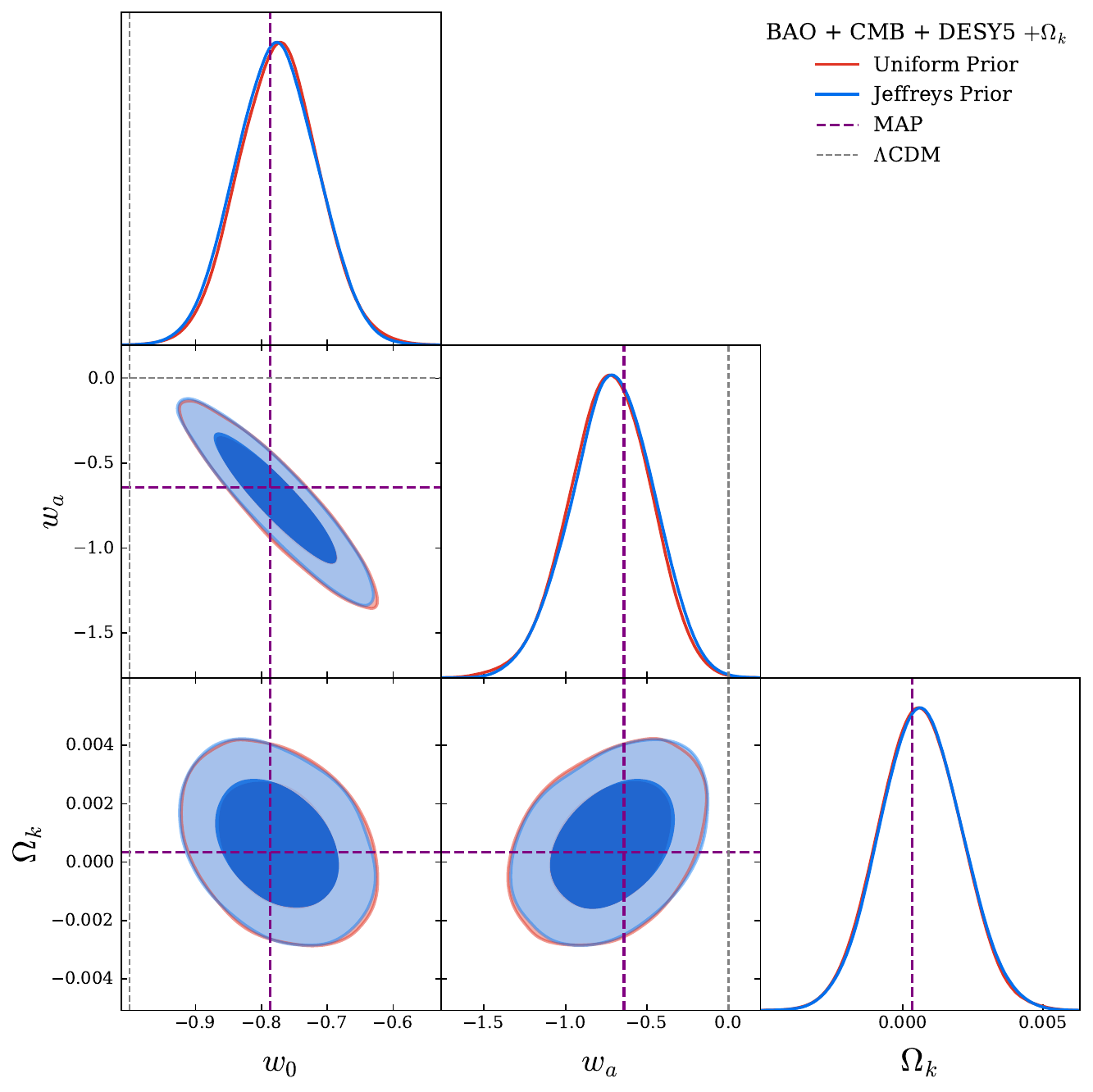}
    \includegraphics[width=0.495\linewidth]{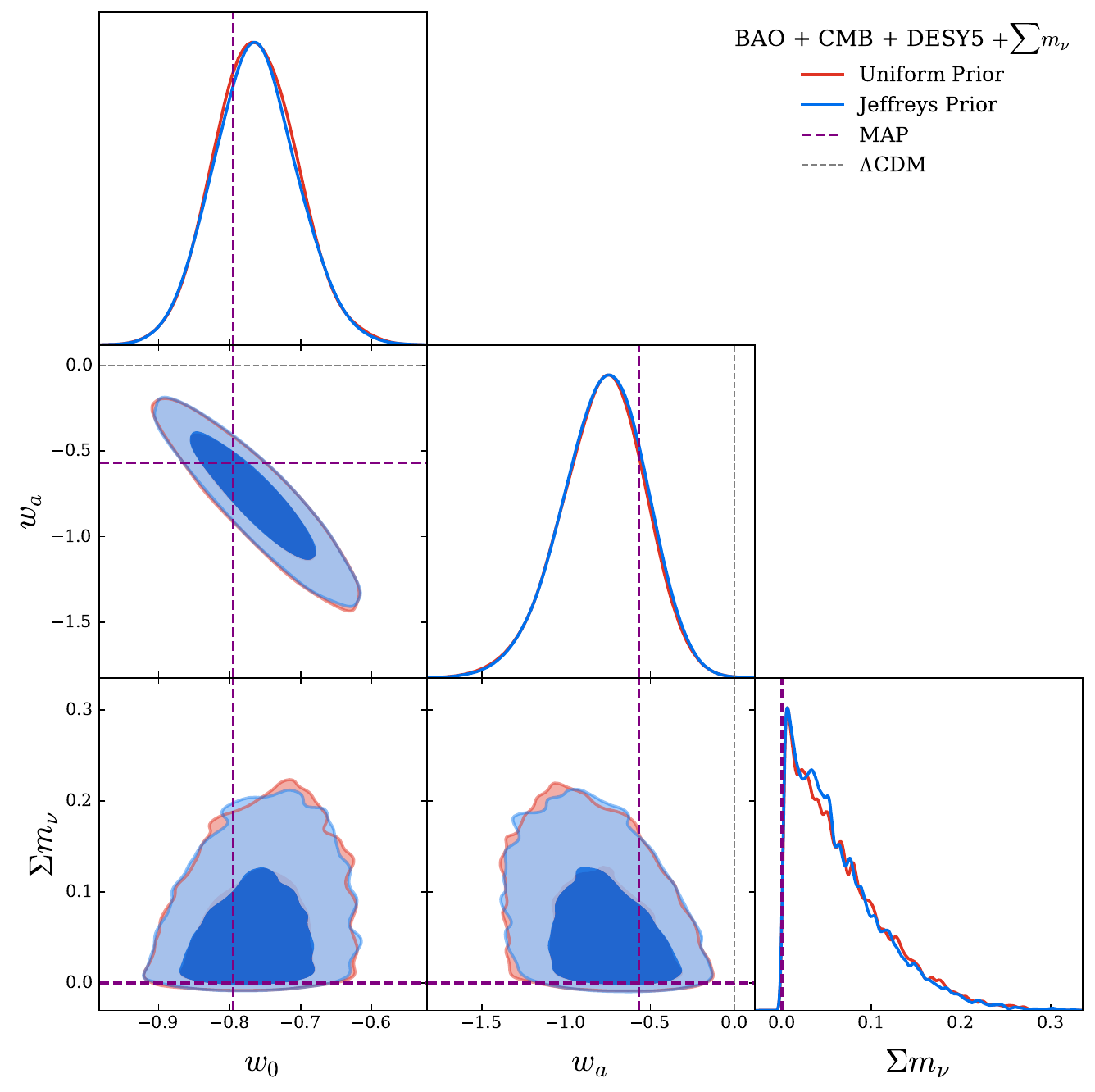}
    \caption{Plots of the \(w_0-w_a\) marginalised posterior with uniform priors (red) and a Jeffreys prior (blue) with free curvature (top) and neutrino mass (bottom). The purple dashed line is the \textit{Maximum A Posteriori} (MAP) estimate computed using the uniform prior posterior. These extensions are used to investigate a larger parameter space and, in the case of neutrino mass, increased degeneracy.}
    \label{fig:Extensions}
\end{figure}

Extensions to $w_0w_a$CDM are explored to increase the influence of the prior on the posterior. In the Left of Figure \ref{fig:Extensions}, we compare the posteriors obtained when allowing free curvature. The same behaviour as in flat \(w_0w_a\)CDM is observed in this extended parameter space, where no significant shifts occur. The same narrative continues when reviewing the right of the figure where neutrino mass is allowed to vary. The extended parameter space and increased degeneracy volume introduced by free neutrino mass gives rise to increased prior volume effects, which we see by a large displacement of the uniform marginalised posterior away from the \textit{maximum m posteriori} (MAP) estimate. The Jeffreys prior does not indicate any signs of alleviating this discrepancy, nor does it show any movement in the neutrino mass posterior towards more physical values. The overall lack of minimising any discrepancy between the marginalised posterior modes and the MAP, especially in highly degenerate cases, highlights that the Jeffreys prior is not designed to remove projection effects, it only tends to mitigate them; hence, it is not surprising that we don't observe movement towards the MAP.
\FloatBarrier

\end{document}